\documentclass[prb,aps,twocolumn,amsmath,amssymb,showpacs]{revtex4}

\usepackage{graphicx}
\usepackage{subfigure}
 
\begin{document}

%
% Title Page
%

\title{On Interband Pairing in Multiorbital Systems}

\author{Adriana Moreo, Maria Daghofer, Andrew Nicholson, and Elbio Dagotto}
 
\affiliation{Department of Physics and Astronomy,University of Tennessee,
Knoxville, TN 37966-1200}
\affiliation{Oak Ridge National Laboratory,Oak Ridge,
TN 37831-6032,USA, }

\date{\today}

\begin{abstract}

The discovery of high-$T_c$ superconductivity in the pnictides, materials with a Fermi 
surface determined by several bands, highlights the need to 
understand how superconductivity arises in multiband systems. In this effort,
using symmetry considerations and mean-field approximations, 
we discuss how strong hybridization among orbitals may lead to both 
intra and 
interband pairing, and we present calculations of the spectral functions to 
guide the experimental search for this kind of state.

\end{abstract}
 
\pacs{74.20.-z, 74.20.De, 74.20.Rp}
 
\maketitle

\section{Introduction}

Iron-based high-$T_c$ 
superconductors\cite{Kamihara:2008p932,chen1,chen2,wen,chen3,ren1,55,ren2} have a complex
Fermi surface that is determined by several bands, an effect resulting from the hybridization
of the $3d$ 
orbitals of iron.\cite{first,Singh:2008p1736,xu,cao,fang2} Band structure calculations
have shown that the bands that define the two hole pockets around the 
$\Gamma$ point have mostly $d_{ xz}$ and $d_{yz}$ character, while the two
electron pockets around the M point have $d_{xz}$, $d_{yz}$, and a smaller amount 
of $d_{xy}$ contributions.\cite{plee,first,Singh:2008p1736,xu,cao,fang2} For this 
reason it is 
important to understand superconductivity in multiorbital systems in general terms. 

Among
the first to address this complex problem 
several years ago were Suhl {\it et al.}\cite{Suhl} using a model consisting of
two orbitals, one $s$ and one $d$, that did not hybridize with each other. Thus, 
in this case each band was determined by one single orbital. They showed that
BCS pairing\cite{BCS} could occur in each band and, since in the most general 
case the
electron-phonon interaction would have different strengths for electrons in 
the different bands, it was proposed that two different superconducting gaps
could arise. Almost 50 years were needed to observe experimental evidence of this 
phenomenon. In 2001, superconductivity with $T_{c}=39$~K was observed in 
MgB$_2$.\cite{akimitsu} Despite the high-$T_{c}$, 
it became clear that
the BCS mechanism\cite{Louie} was at play and for the first time two different 
superconducting gaps were observed.\cite{Wang2001179,PhysRevLett.87.047001,PhysRevLett.87.167003,PhysRevLett.87.137005,PhysRevLett.87.177008,PhysRevLett.87.157002,PhysRevLett.87.177006} As shown in 
Ref.~\onlinecite{Louie}, the Fermi surface (FS) is determined by two bands: the
$\pi$ band formed by the $p_z$ orbitals of B, and the $\sigma$ band constituted
by a linear combination of the $p_x$ and $p_y$ B-orbitals. 
Although three orbitals determine the FS, 
it is interesting to notice that only two different BCS
gaps are observed. This occurs because two of the three orbitals  
hybridize with each other and determine one single band, which couples 
strongly to the lattice phonons. This opens a large superconducting gap 
on the $\sigma$ FS. The other orbital, $p_z$,
does not hybridize and forms the $\pi$ band that couples weakly to the lattice
phonons determining a second, smaller, superconducting gap at the FS of the 
$\pi$ band. Thus, the number of different gaps that can arise in a
multiorbital system is related to the degree of hybridization among
the orbitals. Also note  
that in this early effort {\it interband hopping of pairs of electrons belonging to the same band was included} but the possibility of 
interband pairing, i.e., {\it pairs formed by electrons belonging to two different bands}, was not considered. 

In this paper, the subject of superconductivity 
in multiorbital systems is revisited, in particular to shed light on the possible symmetry 
of the pairing operator of the pnictides superconductors. 
The motivation is that the
pairing operators that have been discussed the most thus 
far\cite{kuroki,Mazin:2008p1695,FCZhang,han,korshunov,Baskaran:2008p832,plee,yildirim,Si:2008p1561,yao,xu2,stratos} assume 
that only intraband pairing should occur, namely the two electrons of the Cooper pair belong to the same band.\cite{dolgov} 
However, numerical simulations\cite{ours} performed on a 
two-orbital model\cite{scalapino,ours,moreo} for the pnictides favor an 
interorbital pairing operator that, when transformed to the band 
representation, results not only in intraband 
pairing but it includes interband pairing as well.\cite{moreo}
For the pnictides, interband hopping of pairs formed between electrons
in {\it the same} band is often denoted as ``Interband Superconductivity''.\cite{dolgov} 
The situation discussed in the present paper is {\it different} and involves
Cooper pairs where the two electrons come from two different bands,
which we will call ``Interband pairing''. Using symmetry arguments and
mean-field approximations, the plausibility and physical meaning of
such an interband pairing in multiorbital systems will be discussed. 

\begin{figure}[thbp]
\begin{center}
\includegraphics[width=8cm,clip,angle=0]{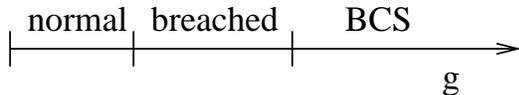}
\vskip 0.3cm
\caption{Schematic representation of the three regimes that can arise
as a function of the strength $g$ of an interband pairing attraction. The label 
``normal'' denotes
a non-superconducting state. The case ``breached'' is an exotic regime with superconductivity
and gaps, coexisting with Fermi surfaces (or several nodes) 
and electrons that do not pair. ``BCS'' is the
large attraction region, where the ground state resembles that of a BCS superconductor
and all electrons participate in the pairing.}
\label{regimes}
\end{center}
\end{figure}

Interband pairing has previously been addressed in the 
context of 
Quantum Chromodynamics (QCD) and cold atoms,\cite{PhysRevLett.90.047002,PhysRevLett.91.032001} 
heavy fermions,\cite{khomskii} cuprates\cite{kheli}, 
and BCS superconductivity.\cite{kumar} In the case of heavy fermions, it was 
argued that interband pairing could occur if two Fermi surfaces arising from different bands are 
very close to one other,\cite{khomskii} while in QCD and cold atoms it was 
presented as a possibility for the case of sufficiently strong attractive pairing 
interactions, or for weaker attractions among particles with very different
masses.\cite{PhysRevLett.90.047002,PhysRevLett.91.032001} As it will be 
discussed for a simple model in Sec.~\ref{sec:interband},
three different regimes, shown 
schematically in Fig.~\ref{regimes}, can result from a purely interband pairing as a 
function of the strength of the pairing potential $g$: (1) a normal regime where the
ground state is not superconducting (namely in purely interband pairing an 
infinitesimal attraction does not lead to superconductivity), (2) an exotic
superconducting ``breached'' regime where gaps open at the normal Fermi surfaces while 
new Fermi surfaces defining regions containing unpaired electrons are created, and 
(3) a superconducting regime resembling BCS states, at large attractive coupling.\cite{rong}

The paper is organized as follows. 
In Sec.~\ref{sec:pairings},
the general form of pairing operators in multiorbital systems will be presented, 
remarking how the 
symmetry is determined by the spatial and the orbital characteristics of the 
operator. The interorbital pairing operator with $B_{2g}$ symmetry obtained
numerically in a two-orbital model for the pnictides is discussed, emphasizing 
that this operator presents a mixture of intra and interband pairing in the band 
representation. In Sec.~\ref{sec:interband}, a simple toy model with pure interband pairing 
attraction is introduced. This simplified model is discussed in order to 
illustrate the effects of interband pairing
on observables, such as the occupation number and the spectral functions 
$A({\bf k},\omega)$. The stability of the interband paired state is also discussed. 
The occupation number, spectral functions, and the stability of the $B_{2g}$ 
pairing state are the subject of Sec.~\ref{sec:interorbital} which is 
directly related to the physics of pnictides, while
Sec.~\ref{sec:conclusions} is devoted to the  conclusions.

\section{Pairing Operators in Multiorbital Models}\label{sec:pairings}

In single-orbital models, the symmetry of a spin-singlet pairing state
is completely determined by the properties of its spatial form factor. More 
specifically, the pairing operator will have the form:
\begin{equation}
\Delta({\bf k})=f({\bf k)}(c_{{\bf k},\uparrow}c_{{\bf -k},\downarrow}-c_{{\bf k},\downarrow}c_{{\bf -k},\uparrow}),
\label{1}
\end{equation}
\noindent where $c_{{\bf k},\sigma}$ destroys an electron with momentum 
${\bf k}$ and spin projection $\sigma$ and $f({\bf k})$ is the form factor 
that transforms
according to one of the irreducible representations of the crystal's 
symmetry group. Thus, $f({\bf k})$
determines the symmetry of the operator. These form factors depend on the 
lattice geometry and generally they may be very complex. However, in materials with 
short pair coherence lengths, such as the high-$T_c$ cuprates, 
the assumption that the two particles that form 
the pair can be very close to one other is usually made.
The Cu-oxide planes in the cuprates 
have the symmetry properties of the $D_{4h}$ 
group and the case $f({\bf k})=
\cos k_x-\cos k_y$, which transforms according to the irreducible 
representation $B_{1g}$, 
provides the well-known $d$-wave symmetry pairing.  

In multiorbital systems, on the other hand, a spin singlet pairing operator 
will have both spatial and orbital degrees of freedom and it will be given by
\begin{equation}
\Delta({\bf k})=f({\bf k)}
\tau_{\alpha,\beta}
(d_{{\bf k},\alpha,\uparrow} d_{{\bf -k},\beta,\downarrow}
-d_{{\bf k},\alpha,\downarrow} d_{{\bf -k},\beta,\uparrow}),
\label{2}
\end{equation}
\noindent where $d_{{\bf k},\alpha,\sigma}$ destroys an electron with momentum 
${\bf k}$, in orbital $\alpha$, and with spin projection $\sigma$, $f({\bf k})$ is the 
spatial form factor as indicated above, and $\tau_{\alpha,\beta}$ is a
matrix in the space spanned by the orbitals involved.
The dimension of $\tau$ is equal
to the number of orbitals that are considered to be of relevance. In this case, 
notice that the
symmetry of the pairing operator would in general be determined by the product 
of the symmetry properties of $f({\bf k)}$ and the symmetry of the
orbital contribution $\tau_{\alpha,\beta}$.
Only if $\tau_{\alpha,\beta}$ is the identity matrix 
do the orbital contribution become trivial, because the identity
matrix transforms according to $A_{1g}$.
Thus, this is the only case where $f({\bf k})$ fully determines the symmetry of the 
pairing operator, as in the single-orbital example.

The minimum model for the pnictides considers the two orbitals $d_{xz}$ and
$d_{yz}$, which are strongly hybridized.\cite{ours,scalapino} All the possible
pairing operators, up to nearest-neighbor distance, that are allowed by the lattice 
and orbital symmetries have been already 
calculated.\cite{wang,shi,2orbitals,wan,zhou,goswami}
Numerical simulations performed on the two-orbital model suggest that the favored 
pairing operator at intermediate couplings, where the state is both 
magnetic and metallic,~\cite{ours,moreo} has
symmetry $B_{2g}$ and is given by Eq.~(\ref{2}) with 
$f({\bf k)}=(\cos k_x+\cos k_y)$, which transforms according to $A_{1g}$, and
$\tau=\sigma_1$ which transforms according to $B_{2g}$~\cite{wan} 
(where $\sigma_i$ are Pauli matrices). Thus, the non-trivial symmetry 
under rotations arises 
from the orbital portion of the operator. This pairing operator has been studied at the 
mean-field level in Ref.~\onlinecite{moreo}. In the orbital representation, the 
Bogoliubov-de Gennes Hamiltonian matrix is given by:
\begin{equation}\label{eq:bcs_ours}
H_{\rm MF}=
 \left(\begin{array}{cccc}
\xi_{xx}  & \xi_{xy}               &     0         & \Delta_{\bf k} \\
\xi_{xy}  & \xi_{yy}               &     \Delta_{\bf k}  & 0 \\
0         & \Delta_{\bf k}               & -\xi_{xx}     & -\xi_{xy}  \\
\Delta_{\bf k}  & 0                      & -\xi_{xy}     & -\xi_{yy}
\end{array} \right),
\end{equation}
\noindent with

\begin{eqnarray}
\xi_{xx}&=&-2t_2\cos k_x-2t_1\cos k_y-4t_3\cos k_x\cos k_y-\mu, \nonumber \\
\xi_{yy}&=&-2t_1\cos k_x-2t_2\cos k_y-4t_3\cos k_x\cos k_y-\mu, \nonumber \\
\xi_{xy}&=&-4t_4 \sin k_x \sin k_y,
\label{4}
\end{eqnarray}
\noindent and
\begin{equation}
\Delta_{\bf k}=V(\cos k_x+\cos k_y),
\label{5}
\end{equation}
\noindent where $V=V_0\Delta$, with $V_0$ being the strength of the pairing 
interaction and $\Delta$ the mean-field parameter obtained by minimizing
the energy. Since the two
orbitals are hybridized via $\xi_{xy}$, in the band representation
the Hamiltonian matrix becomes:
\begin{equation}
H'_{\rm MF}=
\left(\begin{array}{cccc}
\epsilon_1 & 0               &     V_{B}  & V_{A} \\
0 & \epsilon_2               &     -V_{A}  & V_{B} \\
V_{B}  & -V_{A} & -\epsilon_2 & 0  \\
V_{A}  & V_{B} & 0 & -\epsilon_1
\end{array} \right),
\label{6}
\end{equation}
\noindent where $V_{A}$ and $V_{B}$ are given by
\begin{equation}
V_{A}=2u({\bf k})v({\bf k})\Delta_{\bf k}, \\
\label{7}
\end{equation}
\begin{equation}
V_{B}=(v({\bf k})^2-u({\bf k})^2)\Delta_{\bf k},
\label{8}
\end{equation}
\noindent and $u({\bf k})$ and $v({\bf k})$ 
are the elements of the change of basis matrix $U$
given by
\begin{equation}
U=
 \left(\begin{array}{cccc}
u({\bf k}) & v({\bf k})               &     0  & 0 \\
v({\bf k}) & -u({\bf k})               &     0  & 0 \\
0  & 0 & v({\bf k}) & u({\bf k})  \\
0  & 0 & -u({\bf k}) & v({\bf k})
\end{array} \right) ,
\label{9}
\end{equation}
\noindent with $U^{-1}=U^T$.
Remember that $V_{A}$ and $V_{B}$, 
are functions of the momentum ${\bf k}$, 
and $u({\bf k})^2+v({\bf k})^2=1$. Thus, it is clear that in the band 
representation, in addition to the intraband pairing given by $V_{A}$, there is also
interband pairing given by $V_{B}$. 
Among pairing operators compatible with the symmetry of the model, the
ones that do not lead to interband pairing not only cannot mix
orbitals, but have to contain $\tau=\sigma_0$, i.e., the identity 
matrix.\cite{moreo} One example for such an operator would be the
$s_\pm$ pairing.\cite{kuroki,Mazin:2008p1695,korshunov,parker} 

The discussion above describes general properties of hybridized multi-orbital systems. 
If the orbitals are hybridized, but not related to one another by 
symmetry, there is no reason to expect that the coupling between the electrons
in each orbital, and the interaction that produces the pairing, will have 
the same strength for all the orbitals and lead to a unit matrix in the orbital
sector. Then, it is expected that interband pairing will arise in
general and, thus, it is important to understand its consequences, 
providing the main motivation for the present manuscript.

\section{Interband Pairing}\label{sec:interband}

\subsection{Generic properties}

\subsubsection{Model and non-interacting limit}

\begin{figure}[thbp]
\begin{center}
\includegraphics[width=8cm,clip,angle=0]{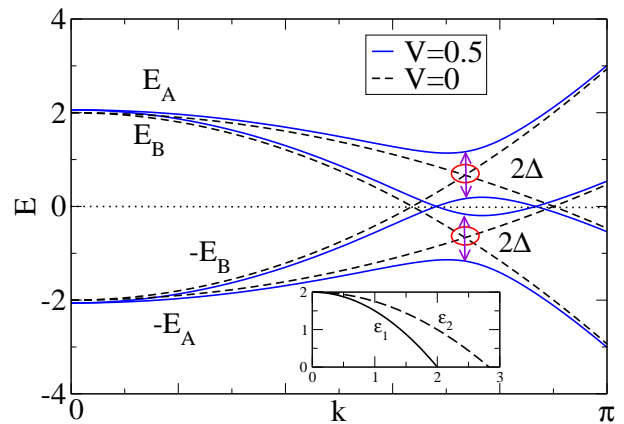}
\vskip 0.3cm
\caption{(Color online) Mean-field band dispersion for the model defined by
Eq.~(\ref{10}), for the indicated values of $V$ (defined in the text) as a
function of the momentum $k=\sqrt{k_x^2+k_y^2}$. The case shown is for 
$m_1=1$, $m_2=2$, and $C=2$. Inset:
Non-interacting band dispersion for the same parameters. }
\label{bands}
\end{center}
\end{figure}

To address qualitatively the issue of 
interband pairing, postponing the matters of stability 
to the next subsection (Sec.~\ref{sec:stable}), 
let us consider the following two-bands simplified model with interband pairing:
\begin{equation}
H_{\bf k}=\sum_{\alpha,\sigma}\epsilon_{\alpha}({\bf k})c^{\dagger}_{{\bf k},\alpha,\sigma}
c_{{\bf k},\alpha,\sigma}+V\sum_{\alpha \ne \beta}
(c^{\dagger}_{{\bf k},\alpha,\uparrow}c^{\dagger}_{-{\bf k},\beta,\downarrow} + h.c.),
\label{10}
\end{equation}
where $\alpha,\beta, = 1, 2$ label two bands that are not hybridized, 
$\sigma$ is the spin projection, and for simplicity
\begin{equation}
\epsilon_{\alpha}({\bf k})={-{\bf k}^2\over{2m_{\alpha}}}+C,
\label{11}
\end{equation}
which gives parabolic bands that are degenerate at ${\bf k}=0$ 
with energy $C$, 
and with a chemical potential $\mu=0$. This can be considered as a crude representation
of the two hole-pocket bands around the $\Gamma$ point in the
pnictides, but more importantly presents a simple toy model where the
effects of interband pairing can be studied. 
As before, the parameter $V=V_0 \Delta$ is the product of 
an attractive potential $V_0$ between electrons in the two different bands 
and a
mean-field parameter $\Delta$ determined by minimizing the total energy. 
The band
dispersion without the interaction is presented in the inset of 
Fig.~\ref{bands}.
The Bogoliubov-de Gennes matrix expressed in the basis expanded by 
$B_{12}=\{c^{\dagger}_{{\bf k},1,\uparrow},c_{-{\bf k},2,\downarrow},c^{\dagger}_{{\bf k},2,\uparrow},c_{-{\bf k},1,\downarrow}\}$ has the form:
\begin{equation}
H=
\left(\begin{array}{cccc}
\epsilon_1 & V               &    0  & 0 \\
V & -\epsilon_2               &  0  & 0 \\
0  & 0 & \epsilon_2 & V  \\
0  & 0 & V & -\epsilon_1
\end{array} \right).
\label{eq:12}
\end{equation}
This matrix can be diagonalized becoming
\begin{equation}
H_D=
\left(\begin{array}{cccc}
E_{A} & 0               &    0  & 0 \\
0 & -E_{B}               &  0  & 0 \\
0  & 0 & E_{B} & 0  \\
0  & 0 & 0 & -E_{A}
\end{array} \right),
\label{12a}
\end{equation}
\noindent in the basis 
expanded by $B_{AB}=\{\gamma^{\dagger}_{{\bf k},{\rm A},\uparrow},
\gamma_{-{\bf k},{\rm B},\downarrow},\gamma^{\dagger}_{{\bf k},{\rm B},\uparrow},
\gamma_{-{\bf k},{\rm A},\downarrow}\}$ and the change-of-basis matrix is given by
\begin{equation}
U=
\left(\begin{array}{cccc}
u_{\bf k} & v_{\bf k}               &    0  & 0 \\
-v_{\bf k} & u_{\bf k}               &  0  & 0 \\
0  & 0 & u_{\bf k} & v_{\bf k}  \\
0  & 0 & -v_{\bf k} & u_{\bf k}
\end{array} \right),
\label{12b}
\end{equation}
\noindent with $u_{\bf k}^2+v_{\bf k}^2=1$.

\begin{figure}[thbp]
\begin{center}
\includegraphics[width=8cm,clip,angle=0]{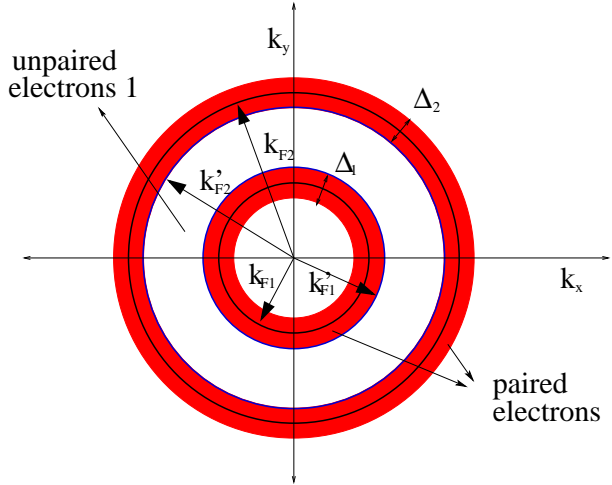}
\vskip 0.3cm
\caption{(Color online) Schematic diagram of the FS determined by the
two parabolic bands of the simple model used in
Sec.~\ref{sec:interband}. $k_{F1}$ and $k_{F2}$ indicate the
Fermi momentum of the two bands when $V=0$, while $k'_{F1}$ and $k'_{F2}$
indicate the position of the Fermi momenta for the case of a finite but small pairing
potential. The shaded rings indicate the regions with width $\Delta_1$ and 
$\Delta_2$ in momentum space where 
electrons can pair. The white region in between the rings contains unpaired 
electrons in band 1.}
\label{diag}
\end{center}
\end{figure}

The four energy eigenvalues are
\begin{equation}
E_i({\bf k})=\pm\left[{(\epsilon_1-\epsilon_2)\over{2}}
\pm\sqrt{({\epsilon_1+\epsilon_2\over{2}})^2+V^2}\right],
\label{13}
\end{equation}
where the first positive (negative) sign corresponds to the
eigenvalues of the upper (lower) block, labeled $E_{A}$ and
$-E_{B}$ ($E_{B}$ and $-E_{A}$) in Eq.~(\ref{12a}) and in
Fig.~\ref{bands}. The second sign differentiates between the two
solutions in each block.
Figure~\ref{bands} shows the eigenvalues for $V=0$ and $V=0.5$. When $V=0$,
then $E_{A}=\epsilon_2$ and $E_{B}=\epsilon_1$, the two bands
define two circular Fermi surfaces with radius $k_{F1}$ and
$k_{F2}$ where they cross the chemical potential ($\mu=0$). This is illustrated 
schematically in Fig.~\ref{diag}. 

\begin{figure}[thbp]
\begin{center}
\includegraphics[width=8cm,clip,angle=0]{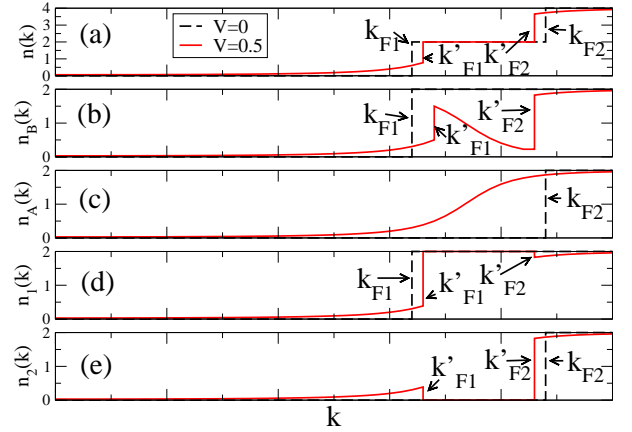}
\vskip 0.3cm
\caption{(Color online) Mean-field state population as a function of
momentum along the diagonal $k_x=k_y$ 
for (a) the whole system; (b) band B, (c) band A; (d) orbital 1; and (e)
orbital 2 for the indicated 
values of the pairing potential $V$, and $m_1=1$, $m_2=2$, and $C=2$.}
\label{nkab12}
\end{center}
\end{figure}

Notice that number operators can be defined in the two bases that are being considered
here, i.e. $B_{12}$ and $B_{AB}$, which of course 
become equivalent when $V$=0. Thus, in the
basis $B_{12}$ the number operator is,
\begin{equation}
n_{\alpha}({\bf k})=\sum_{\sigma}c^{\dagger}_{{\bf k},\alpha,\sigma}c_{{\bf k},\alpha,\sigma},
\label{12c}
\end{equation}
\noindent while in 
basis $B_{AB}$ the number operator is ($I$=A,B)
\begin{equation}
n_{I}({\bf k})=\sum_{\sigma}\gamma^{\dagger}_{{\bf k},{\rm I},\sigma}\gamma_{{\bf k},{\rm I},\sigma}.
\label{12d}
\end{equation}
The total electronic occupation of the system is given by
\begin{equation}
n({\bf k})=\sum_{\alpha}n_{\alpha}({\bf k})=\sum_{I}n_{I}({\bf k}).
\label{12e}
\end{equation}
Then, for $V=0$ we find that $n({\bf k})=0$ for $|{\bf k}|\le |{\bf k}_{F1}|$, 
$n({\bf k})=2$ for
$|{\bf k}_{F1}|<|{\bf k}|<|{\bf k}_{F2}|$ 
since in this region $n_1({\bf k})=n_{A}({\bf k})=2$ while 
$n_2({\bf k})=n_{\bf B}({\bf k})=0$, 
and finally $n({\bf k})=4$ for $|{\bf k}|>|{\bf k}_{F2}|$ since both orbitals are 
totally filled with electrons. These results are represented by the dashed 
lines in Fig.~\ref{nkab12}.

\subsubsection{Weak attraction}\label{sec:interband_weak}

In the nontrivial case of $V$ different 
from zero, the bands $\pm E_{A}$ and $\pm E_{B}$ result from the hybridization of
the $\epsilon_1$ and $\epsilon_2$ bands due to $V$. It is interesting to observe that an 
internal gap opens at the crossing of bands $E_{A}$ with $-E_{B}$ $above$ the 
chemical potential and between $E_{B}$ and $-E_{A}$ $below$ 
the chemical potential, as
indicated with circles in Fig.~\ref{bands} where results for $V=0.5$ are 
displayed. These are very important differences with respect to conventional BCS
calculations were all the action is restricted to the original Fermi surfaces:
for multiorbital models, gaps can open in other portions of the band structure as well.

While the $\pm E_{A}$ bands are separated by a gap, the
bands $\pm E_{B}$ still cross the chemical potential determining two Fermi surfaces at 
$k'_{F1}>k_{F1}$ and at $k'_{F2}<k_{F2}$, even in this pairing state
(again, this is different from the one-orbital pairing standard BCS ideas).
The new Fermi surfaces are shown 
schematically in Fig.~\ref{diag}: the interior FS has expanded
and the exterior one has contracted. In fact, it will be shown below that 
the effect of $V$ is to try to equalize the two Fermi surfaces, as this pairing attraction grows
in magnitude. Calculating $n({\bf k})$
we observe that $n({\bf k})=4v_{\bf k}^2$ for $k<k'_{F1}$ and $k>k'_{F2}$, 
where $v_{\bf k}$ is the 
element of the change-of-basis matrix in Eq.~(\ref{12b}). This agrees with the 
BCS expression for $n({\bf k})$ but it jumps discontinuously to $n({\bf k})=2$ for 
$k'_{F1}<k<k'_{F2}$. Such a result 
is shown by the solid lines in Fig.~\ref{nkab12}(a). 
The jumps indicate the existence of 
the two  Fermi surfaces, which are here present even in the paired state. 
Thus, some electrons in the region in between the two Fermi surfaces 
may behave like normal unpaired  electrons. 

A better understanding of the electronic behavior can be achieved by
studying the electronic density in the two bases $B_{AB}$ and $B_{12}$. 
We find that $n_{A}({\bf k})=2v_{\bf k}^2$, 
as shown in Fig.~\ref{nkab12}(c), which is the standard BCS 
behavior in agreement with the fact that bands $\pm E_{A}$ are separated by a 
gap. Thus, all the electrons in this band participate in the pairing and they do 
not have a FS. On the other hand, it can be shown 
that $n_{B}({\bf k})=n_{A}({\bf k})=2v_{\bf k}^2$ for
$k<k'_{F1}$ and $k>k'_{F2}$, but for $k'_{F1}<k<k'_{F2}$ there is a 
discontinuous change of behavior to $n_{B}({\bf k})=u^2_{\bf k}=1-v^2_{\bf k}$ 
[Fig.~\ref{nkab12}(b)].
Thus, the Fermi surfaces are 
determined by electrons in the band $E_{B}$. However, note that 
in this region $n_{A}({\bf k})$ is an 
increasing function of $|{\bf k}|$ 
while $n_{B}({\bf k})$ is a decreasing function of $|{\bf k}|$ which
satisfies $n_{A}({\bf k})+n_{B}({\bf k})=2$ for all $|{\bf k}|$. 

This behavior can be better understood by 
calculating the (photoemission) spectral functions $A({\bf k},\omega)$, 
which allow us to obtain 
$n({\bf k})=\int_{-\infty}^{\mu=0}A({\bf k},\omega)d\omega$. 
\begin{figure}[thbp]
\begin{center}
\includegraphics[width=0.4\textwidth,clip,angle=0]{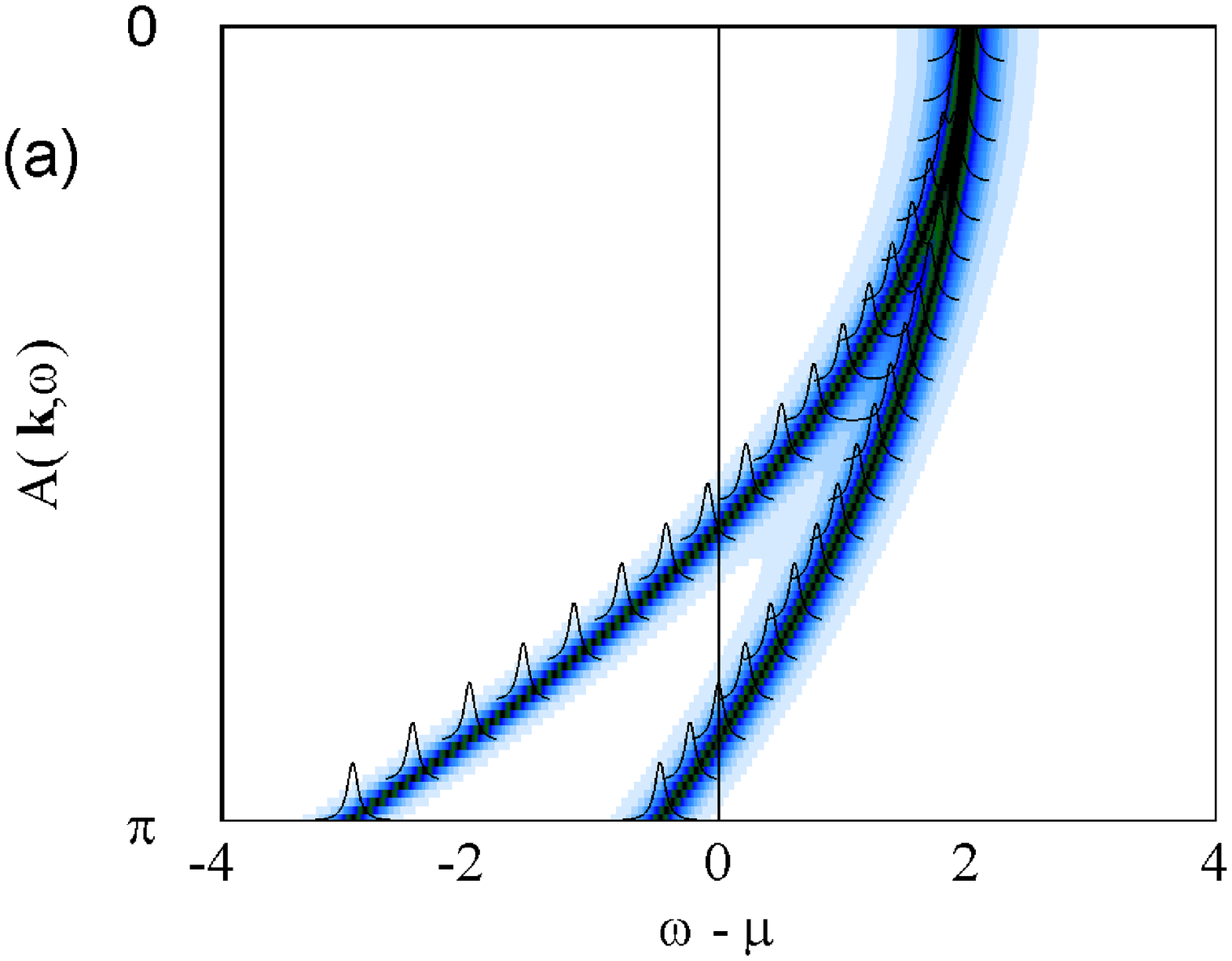}
\includegraphics[width=0.4\textwidth,clip,angle=0]{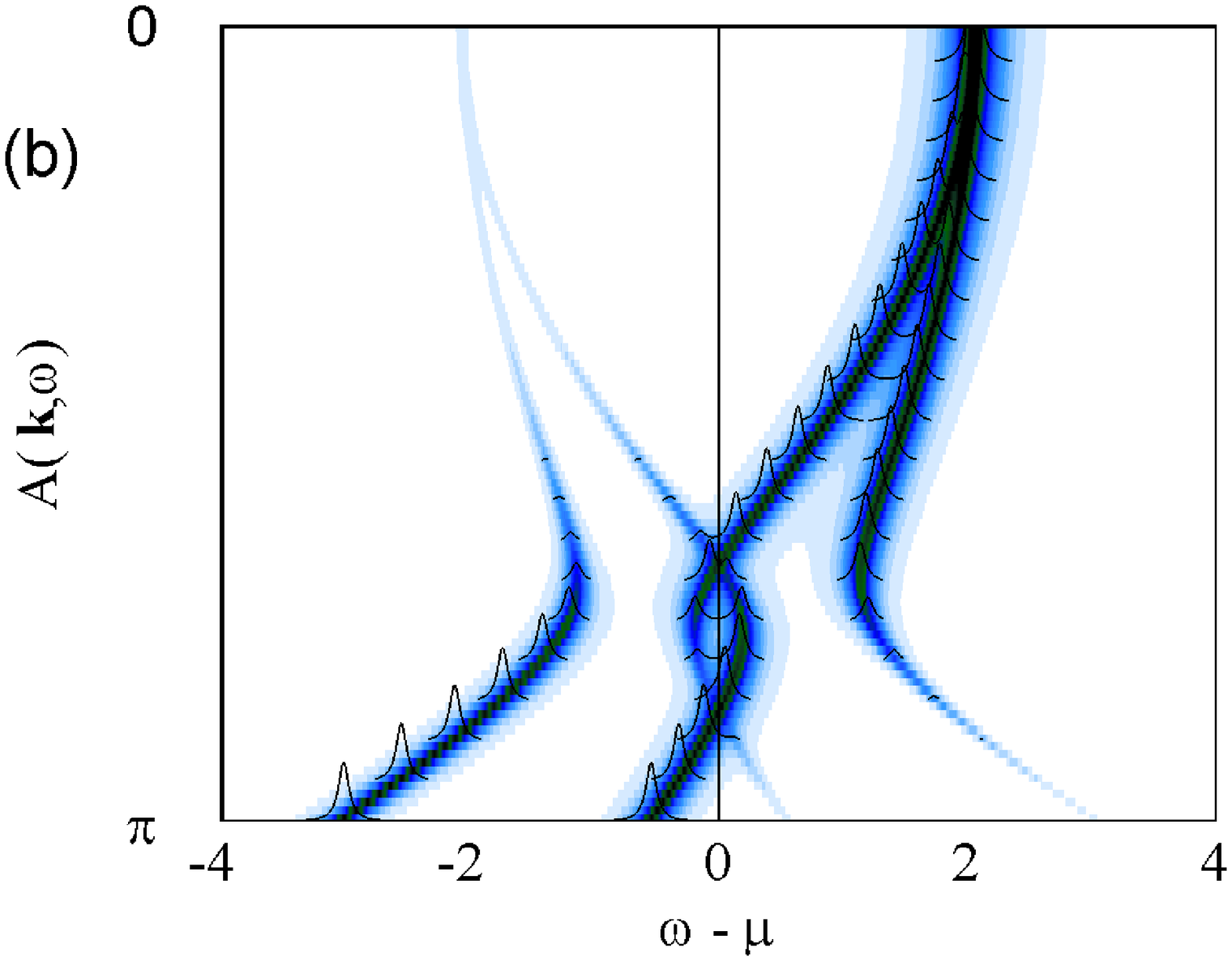}
\includegraphics[width=0.4\textwidth,clip,angle=0]{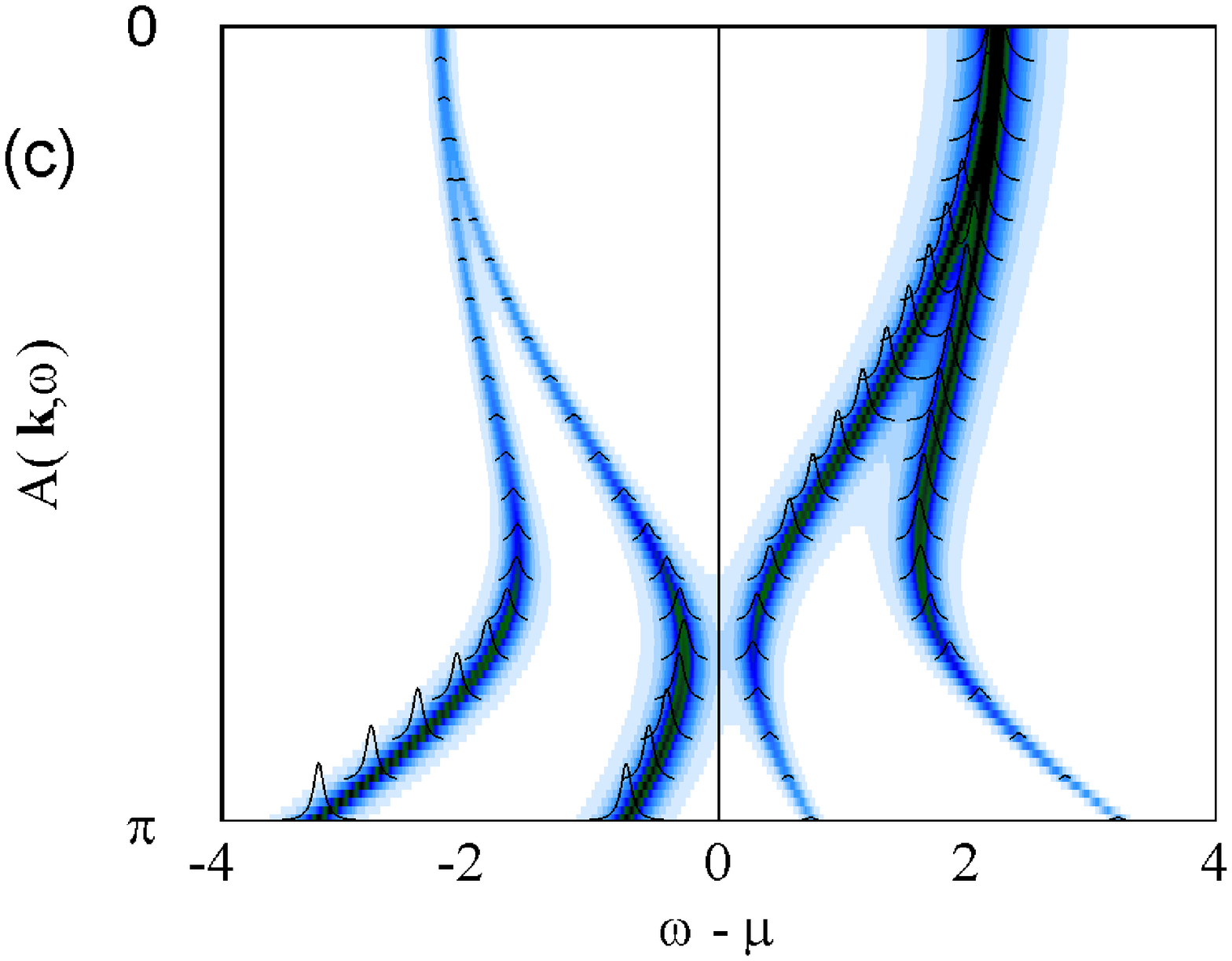}
\vskip 0.3cm
\caption{(Color online) (a) Spectral functions $A({\bf k},\omega)$ for the 
non-interacting case, $V=0$. (b) Same as (a) but for $V=0.5$. Note that there 
are still Fermi surfaces in this intermediate coupling case, namely 
gaps open far from the chemical potential but not at the Fermi level.
(c) Same as (a) but for $V=1.0$.
The presence of a gap for all the momenta shown is now clear.}
\label{akw0}
\end{center}
\end{figure}
In the non-interacting case, shown in Fig.~\ref{akw0}(a), the spectral function
shows two peaks, corresponding to the two bands $\epsilon_1$ and $\epsilon_2$,
for each value of $|{\bf k}|$. This is the expectation for free electrons in a 
non-interacting multiorbital system. The two non-interacting FS's are 
located where 
each band passes across the chemical potential.

When the pairing interaction becomes
finite, the spectral function 
develops four peaks for each value of 
$|{\bf k}|$ as it can be observed in Fig.~\ref{akw0}(b) for $V=0.5$. This is 
also the 
expected result since the BCS interaction generates a ``shadow'' or Bogoliubov 
band for each band present in the non-interacting system. For example, 
close to $|{\bf k}|=0$ the bands $E_{A}$ and $E_{B}$ have almost all 
the spectral weight, given by $u_{\bf k}^2$ which 
is very close to 1 in this region, and follow a dispersion
similar to the non-interacting bands $\epsilon_1$ and $\epsilon_2$, while the 
bands $-E_{A}$ and $-E_{B}$ appear with very small spectral weight 
given by 
$v_{\bf k}^2=1-u^2_{\bf k}$. The latter are the
Bogoliubov or ``shadow'' bands. These shadow bands appear above the chemical 
potential for large values of $|{\bf k}|$, as expected. 

But what happens in the intermediate 
region  $k'_{F1}<k<k'_{F2}$? 
We see that the band $E_B$ crosses with
$-E_B$ at $k'_{F1}$ determining a FS. Most of the spectral weight below 
the chemical potential belongs to the band $-E_B$ which contains the unpaired
electrons but the band $-E_A$ has 
appreciable shadow spectral weight indicating that there are also some paired
electronic population as indicated in Figs.~\ref{nkab12}(b) and (c). As $k$ 
increases, spectral weight is transferred continuously from $-E_B$ to $-E_A$, 
behavior associated with the internal gap opened by the pairing interaction, 
so that when $k$ approaches $k'_{F2}$ most of the spectral weight below 
the chemical potential is in $-E_A$, paired electrons, and $-E_B$, unpaired
electrons, has shadow spectral weight
as seen in Figs.~\ref{nkab12}(b) and (c). At $k=k'_{F2}$ the second FS is 
determined by the crossing of $\pm E_B$ at the chemical potential indicated by
the sudden jump in $n_B$. Thus, the unpaired electrons in this 
region coexist with paired electrons and the spectral functions do not 
resemble the non-interacting ones.

It is also illuminating to analyze what happens with $n({\bf k})$ in the basis $B_{12}$.
While $n_1({\bf k})=n_2({\bf k})=2v^2_{\bf k}$ 
in the regions where there are no unpaired electrons, we
find discontinuities associated with Fermi surfaces in the two distributions 
that are
given by  $n_2({\bf k})=0$ and $n_1({\bf k})=2$ 
for $k'_{F1}<k<k'_{F2}$ [Figs.~\ref{nkab12}(d,e)].
This indicates that the pairing interaction $V$ has been able to promote some 
electrons from above to below $k_{F2}$ in orbital 2. Also, electrons
have been transferred from their original location in orbital 1, in the
neighborhood of $k_{F2}$ and above $k_{F1}$, to both orbitals 1 and
2 around $k_{F1}$.
These are the electrons in 1 and 2 that have become paired (the pairing is 
indicated by the shadowed circular regions in Fig.~\ref{diag}). But the 
interaction was not strong enough to provide pairing partners to all the 
extra electrons originally in orbital 1 and, thus, they have been left unpaired
in between the two paired regions, as indicated in Fig.~\ref{diag}.

Thus, the interband pairing attraction creates pairs of electrons 
belonging to different orbitals within an interval $\Delta_{k_{Fi}}$ around 
each of the two original Fermi surfaces. The width of the pairing region increases with 
$V$. The pairing partners are obtained by promoting electrons with momentum 
$k \approx k_{F2}$ and $k \approx k_{F1}$ in both orbitals and by
moving electrons from the more populated to the less populated orbital.
This creates the conditions to pair electrons near both Fermi surfaces. The 
electrons in band 
1 that could not find promoted partners remain unpaired. Whether this state is 
stable or not depends, of course, on the balance between kinetic and
pairing energies, which will be discussed in Sec.~\ref{sec:stable}.

\subsubsection{Strong attraction}\label{sec:interband_strong}

\begin{figure}[thbp]
\begin{center}
\includegraphics[width=8cm,clip,angle=0]{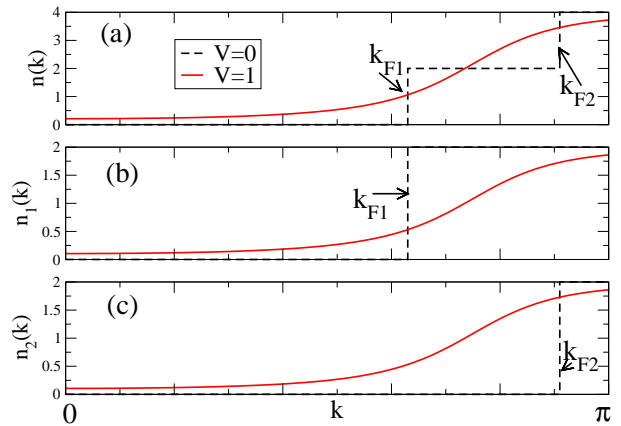}
\vskip 0.3cm
\caption{(Color online) Mean-field state population as a function of the 
momentum (main diagonal) 
for (a) the whole system, (b) band 1, and (c) band 2 for the indicated 
values of the pairing potential $V$, and for $m_1=1$, $m_2=2$, and $C=2$. The case 
$V=1.0$ illustrates the ``strong'' pairing attraction 
regime where gaps open in the original
Fermi surface.}
\label{nk1}
\end{center}
\end{figure}

As the interaction $V$ increases further, the number of unpaired electrons is reduced.
This means that $k'_{F1}$ and $k'_{F2}$ become closer to each other making the
size of the intermediate region with unpaired electrons in Fig.~\ref{diag} 
smaller. Eventually, 
the two momenta become the same $k'_{F1}=k'_{F2}$ for
$V=0.71$, and for $V>0.71$ the region with unpaired electrons vanishes 
and a full gap opens in the system whose physics now resembles BCS, except for 
the fact that the pairs are constituted by electrons from different orbitals. 
The electronic population of the system in such a case, e.g. at $V=1.0$, is 
presented in Fig.~\ref{nk1} and the corresponding spectral functions are shown
in Fig.~\ref{akw0}(c). It can be shown that now $n_1({\bf k})=n_2({\bf k})$ for all values of $k$ and 
all the particles around the two noninteracting Fermi surfaces now participate in the pairing. 
The spectral functions show four peaks, i.e. Bogoliubov bands for all values 
of $k$, as it can be observed in Fig.~\ref{akw0}(c).

Thus, notice that while band 1 contained more electrons than band 
2 for $V=0$ [see Figs.~\ref{nk1}(b) and (c)], the interband pairing mechanism
transfers electrons from one band to the other so that both bands have
the same number of electrons in the superconducting state. Consequently,
the smaller FS expands and the larger one shrinks. Then, when the pairing 
becomes strong enough, the two Fermi surfaces become equalized and no unpaired
electrons remain. Whether this situation can be achieved will depend on the 
strength of the interaction and the energy balance, 
as discussed in the next subsection.

If the two non-interacting Fermi surfaces are very close to each other in momentum space, even
relatively weak pairing interaction could effectively 
be strong enough to make the interband pairing 
resemble BCS pairing as in the case of large $V$ in the present example.

\subsection{Stability of the interband paired state}\label{sec:stable}

\subsubsection{The case without intraband pairing}

As discussed in the Introduction, 
the possibility of interband pairing has been previously discussed in the 
context of QCD and cold atomic matter. Similar 
effects on the FS as found in the present study were 
described, although 
the physics was different because in the QCD context each band contained different 
kinds of particles  and the pairing was thus not able to promote
particles from the majority to the minority band.\cite{PhysRevLett.90.047002,PhysRevLett.91.032001}
The issue of stability was explored in the QCD framework, and it was
found that a purely interband paired state could 
be stabilized for pairing attractions $above$ a certain cut-off value, 
which could become very small for a large difference between the
masses of the two paired species.\cite{PhysRevLett.90.047002,PhysRevLett.91.032001}
 
In the case of our model, however, we have found (see below) 
that the purely interband-paired
state only becomes stable when the attraction is sufficiently strong that no unpaired
particles are left, i.e. when the two shaded regions overlap and the unpaired
region in Fig.~\ref{diag} vanishes. This means that, although the pairs would
be formed by electrons in different orbitals, the physics would be analogous to
BCS. A gap will be opened in the full Fermi surface of the simple model studied here.
 
\begin{figure}
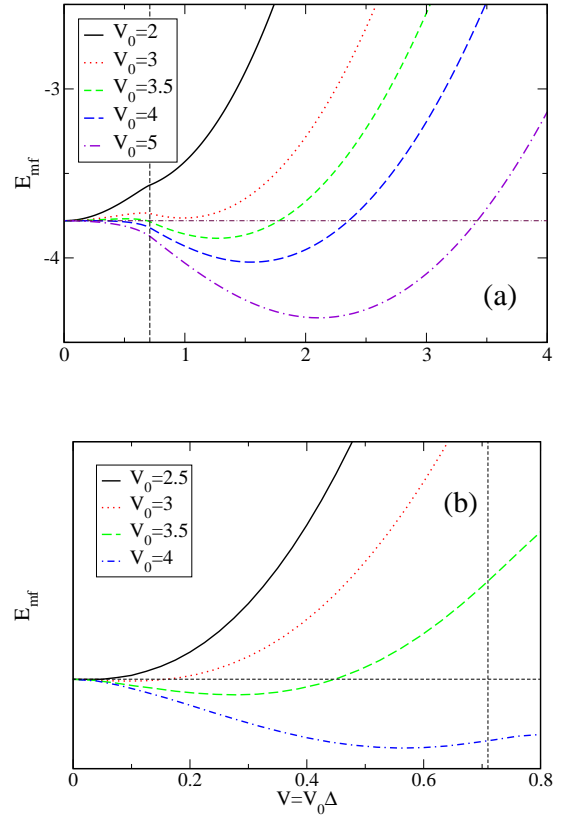

\includegraphics[width=0.4\textwidth]
{paragap_7a}
\vskip 1cm
\includegraphics[width=0.4\textwidth]
{paragapmix_7b}
\caption{(Color online)(a) Mean-field energy Eq.~(\ref{mf}) per site
 vs. $V=V_0\Delta$
for different values of $V_0$. The case $V=0.71$ is 
indicated with dashed lines: it separates
the regions corresponding to nodal and nodeless
states. (b) Mean-field energy (Eq.~\ref{mf}) per site 
with intraband pairing
with strength $V_0/2$ vs. $V=V_0\Delta$,
for different values of $V_0$.\label{fig:stab}}
\end{figure}

In order to study the issue of stability, let us assume that the interaction term responsible
for the interband attraction is given by
\begin{equation}
H_{\rm attr}={1\over{N}}\sum_{{\bf k,k'},\alpha}V_{\bf k,k'}
c^{\dagger}_{{\bf k},\alpha,\uparrow}
 c^{\dagger}_{{\bf -k},-\alpha,\downarrow}c_{{\bf -k'},-\alpha,\downarrow}
c_{{\bf k'},\alpha,\uparrow},
\label{attr}
\end{equation}
\noindent where $V_{\bf k,k'}=-V_0$ and $N$ is the number of sites. 
Performing the standard mean-field
approximation: $b_{\bf k'}=
\langle c_{{\bf -k'},-\alpha,\downarrow}c_{{\bf k'},\alpha,\uparrow} \rangle$ 
and
$b^{\dagger}_{\bf k}=\langle c^{\dagger}_{{\bf k},\alpha,\uparrow}
 c^{\dagger}_{{\bf -k},-\alpha,\downarrow}\rangle$ and making the substitution
$c^{\dagger}_{{\bf k},\alpha,\uparrow} c^{\dagger}_{{\bf -k},-\alpha,\downarrow}=
b^{\dagger}_{\bf k}+(c^{\dagger}_{{\bf k},\alpha,\uparrow}
c^{\dagger}_{{\bf -k},-\alpha,\downarrow}-b^{\dagger}_{\bf k})$
(and an analogous substitution for the product of annihilation
operators), the mean-field results are obtained. As usual, 
the fluctuations around the average given by
$(c^{\dagger}_{{\bf k},\alpha,\uparrow}
c^{\dagger}_{{\bf -k},-\alpha,\downarrow}-b^{\dagger}_{\bf k})$ are assumed to 
be small.
Defining $\Delta={1\over{N}}\sum_{\bf k} b_{\bf k}= {1\over{N}}
\sum_{\bf k} b^{\dagger}_{\bf k}$ we obtain the following mean-field
Hamiltonian:
\begin{eqnarray}
H_{\rm MF}=\sum_{\alpha,\sigma}\epsilon_{\alpha}({\bf k})c^{\dagger}_{{\bf k},\alpha,\sigma}
c_{{\bf k},\alpha,\sigma} \nonumber\\
-V_0\Delta\sum_{{\bf k},\alpha\ne\beta}(c^{\dagger}_{{\bf k},\alpha,\uparrow}c^{\dagger}_{{\bf -k},\beta,\downarrow}+h.c.)+ 2V_0\Delta^2N.
\label{mf}
\end{eqnarray}
Equation~(\ref{eq:12}) can be recovered by defining $-V_0\Delta=V$, 
and disregarding
the constant last term of Eq.~(\ref{mf}). We can calculate the total energy
$E_{\rm MF}$ for
Eq.~(\ref{mf}) as a function of $V=V_0\Delta$. If for a given
$V\ne 0$ the energy has a minimum, this indicates that the interband-paired state is stable. Note that having a term linear in $\Delta$ in Eq.(\ref{mf}) is not
sufficient to conclude the appearance of a superconducting state at small $\Delta$, since the sign of the coefficient of the linear term can change sign with $V$. A similar situation occurs for the magnetic state of undoped pnictides: a finite Hubbard $U$ must be reached to stabilize the ``striped'' state \cite{rong}.
Returning to superconductivity, there
are two regions of interest: (i) $0<V<0.71$, which corresponds to the case in which
two Fermi surfaces are present in the paired state; (ii) $V>0.71$, which corresponds to the
fully gapped case. In Fig.~\ref{fig:stab}(a), $E_{\rm MF}/N$ vs. $V$ is shown 
for different
values of $V_0$. It can be observed that a second minimum develops 
for $V_0>2$
and it becomes stable for $V_0>3$. The minimum always occurs for $V>0.71$
which means that it corresponds to the case in which there are no unpaired
electrons in the system.
The results shown in the figure are robust in the sense that changes in the 
values of $m_1$, $m_2$ or the chemical potential were not found to
stabilize the state with unpaired electrons. Thus, in this respect an attraction
that is only interband can only lead to a stable superconducting state in the
strong attraction region.

\subsubsection{Stability when both inter and intraband pairing coexist}

\begin{figure}[thbp]
\begin{center}
\includegraphics[width=8cm,clip,angle=0]{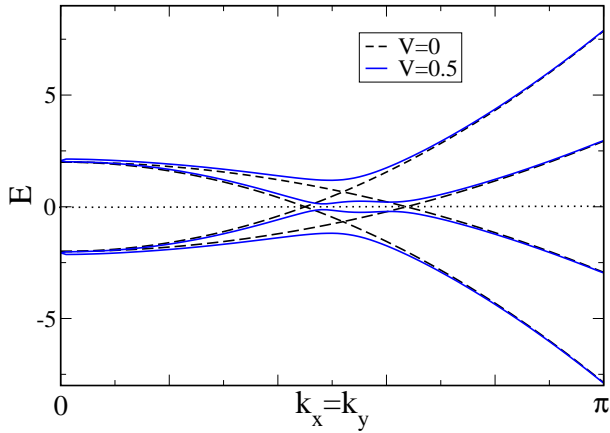}
\vskip 0.3cm
\caption{(Color online) Mean-field band dispersion for the model given by 
Eq.~(\ref{10}) (along the main diagonal) 
for the indicated values of the pairing potential $V$, with 
the addition of an intraband pairing with strength $V_0/2$, as described in 
the text.}
\label{bandasmix}
\end{center}
\end{figure}

The results of the previous paragraphs may seem negative with respect to the
relevance of the ``intermediate'' state with simultaneous coexistence of  
pairing and Fermi surfaces.
However, as pointed out in Sec.~\ref{sec:pairings}, most of the pairing operators allowed by
the lattice and orbital symmetry in the pnictides are characterized by a 
\emph{mixture} of both intra and interband pairing. Thus, 
it is important to consider such a situation 
in our simple model as well. In the case of the $B_{2g}$ pairing operator,
Eqs.~(\ref{6}), (\ref{7}), and (\ref{8}) indicate that the pairing is purely
interband only for $k_x=0$ or $\pi$ 
and $k_y=0$ or $\pi$, because $V_{A}=0$ along these 
lines. Thus, let us now consider our simple model in a Brillouin
zone (BZ) defined by $-\pi<k_x,k_y,\le\pi$ and with the addition of \emph{intraband} 
pairing with intensity $V_0/2$ that is vanishing for $k_x=0$ or $\pi$ and $k_y=0$ 
or $\pi$, i.e.,  the pairing is purely interband only along those directions.
The energy bands now behave as in Fig.~\ref{bands} only along 
$(0,0)-(\pi,0)$ and $(0,0)-(0,\pi)$, while the 
dispersion along any other direction is shown in Fig.~\ref{bandasmix}
for $V=0$ and $V=0.5$.

Performing the mean-field approximation similarly as explained above, we 
have found that the superconducting state now becomes stable on \emph{both} sides
of the original critical value 
$V=0.71$. Figure~\ref{fig:stab}(b) shows that the pairing state is stabilized 
for $V_0>2.5$ but the value of $\Delta$ where the minimum is located is such that
$V<0.71$ and, thus, two nodes will be present along the $x$ and $y$ axes. 
Increasing the value of $V_0$, the minimum eventually occurs for
$V>0.71$. For these larger values of $V_0$, there would consequently be no nodes.
 
Then, in this section it has been shown using a simple model that the interorbital paired
state can become stable if the attraction $V_0$ is sufficiently strong. In this case,
it is the nodeless
case that is stable even for purely interband attraction. In addition, 
the very interesting novel phase with coexisting
nodes and unpaired electrons in the majority band also
requires intraband pairing to be stable, with strength similar to that
of the interband, at least in parts of the BZ.

\section{The interorbital $B_{2g}$ Pairing Operator}\label{sec:interorbital}

In the previous section, a simple model was presented, both with exclusively interband 
pairing and with both inter and intraband pairing, and it was found that the
intraband pairing stabilizes the state with a mixture of superconductivity and metallicity. 
As mentioned in the Introduction, it is expected that in the most  
general cases the pairing operators allowed by the symmetry of the lattice 
and of the orbitals will, in the band representation, have both intra and 
interorbital pairing. Thus, now we will present and discuss, at the mean field
level, the occupation number and the spectral functions for the pairing 
operator obtained from the numerical study of the two-orbital model for the 
superconducting state of the pnictides introduced in Sec.~\ref{sec:pairings}.

\begin{figure}
\subfigure{\includegraphics[width=0.4\textwidth]
{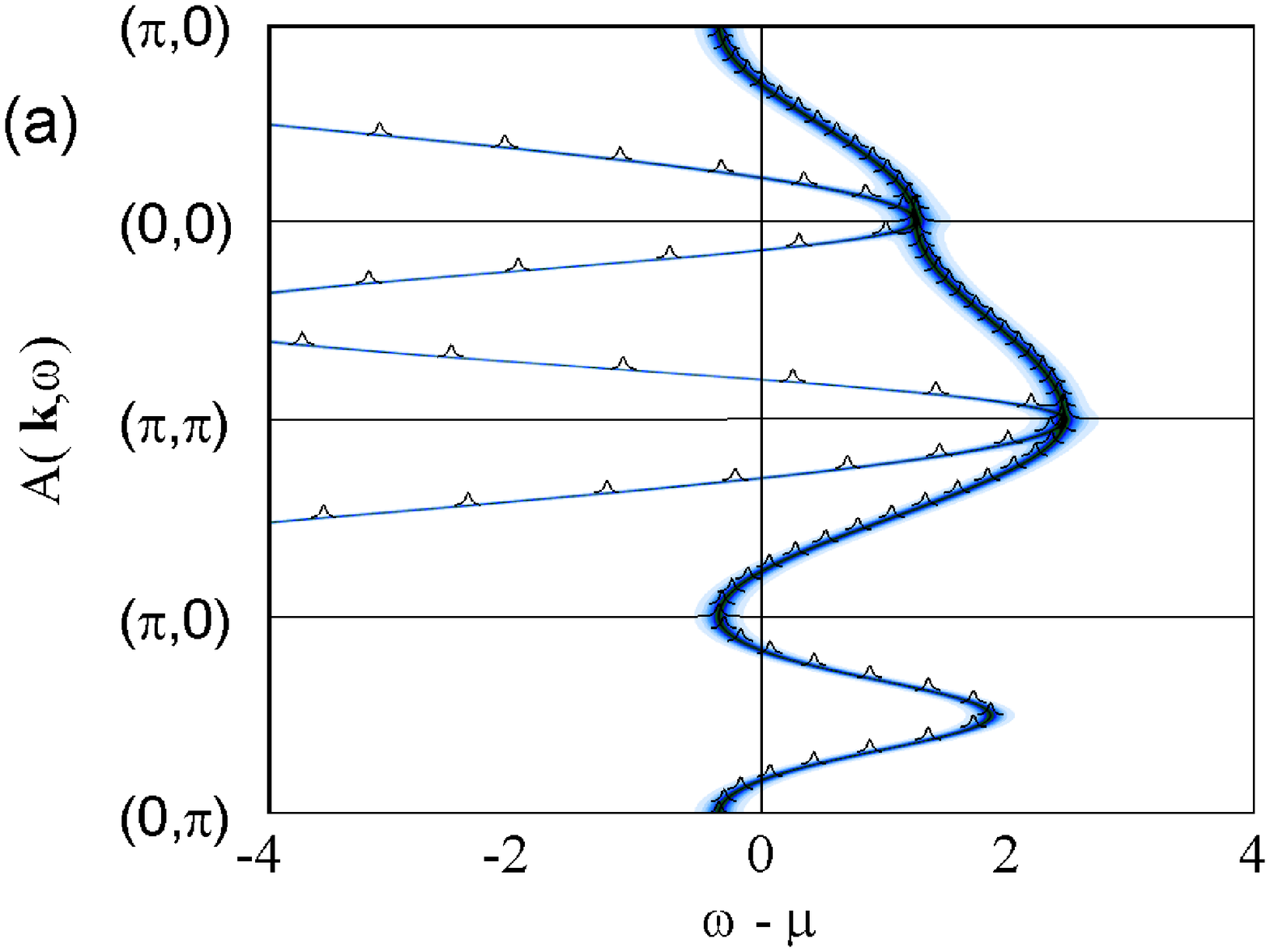}\label{fig:Akb2g_00}}
\subfigure{\includegraphics[width=0.4\textwidth]
{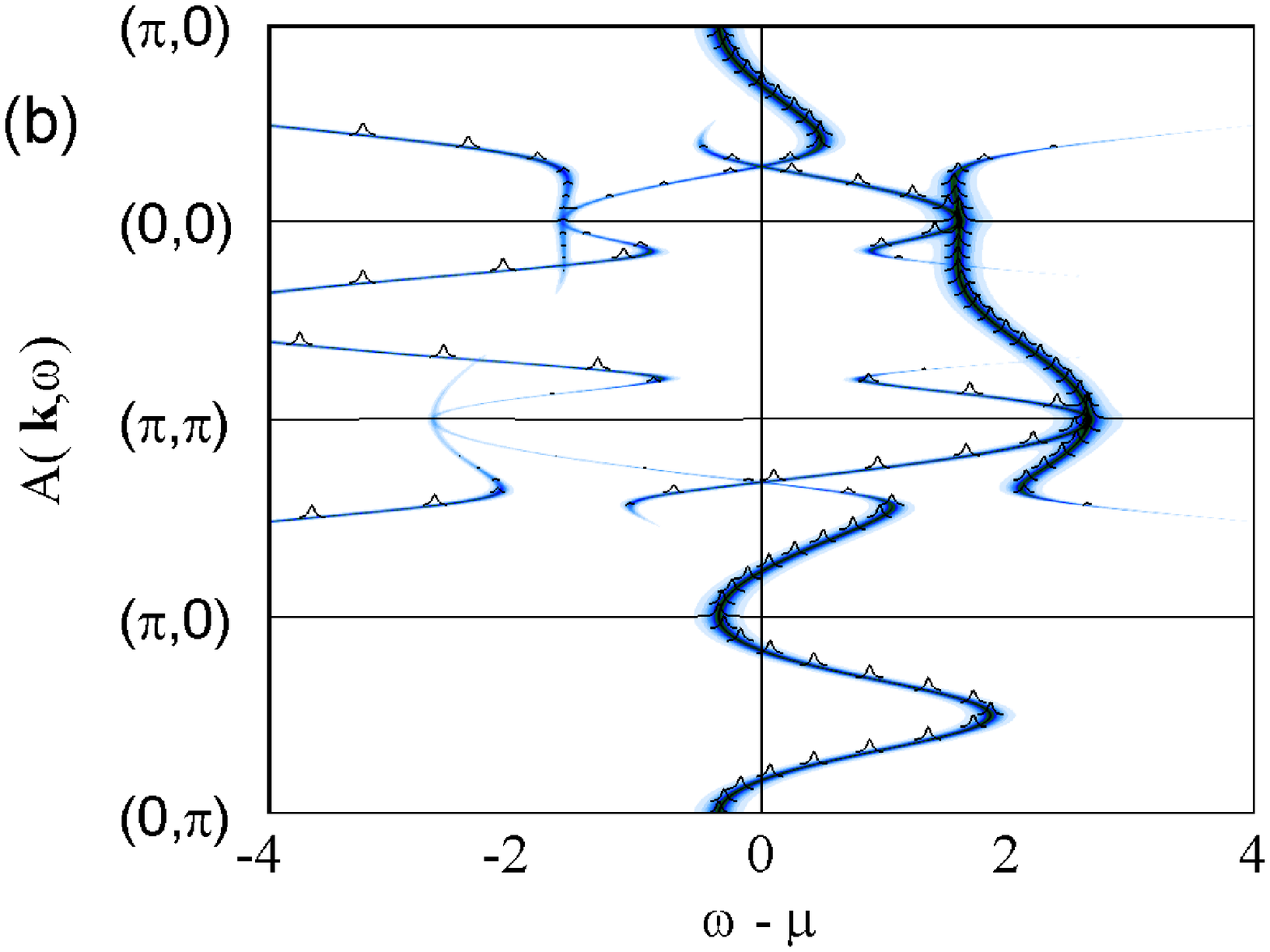}\label{fig:Akb2g_05}}
\caption{(Color online) (a) One-particle spectral function for the 
two-orbital model
  Eq.~(\ref{eq:bcs_ours}),  with vanishing pairing interaction
  $V=0$. Parameters: $t_1=1.3$,  
  $t_2=-1$, $t_3=t_4=-0.85$, $\mu=1.54$. (b) Same as (a) but for $V=0.5$.
\label{fig:Akb2g_0}}
\end{figure}

\begin{figure}
\includegraphics[width=0.4\textwidth]
{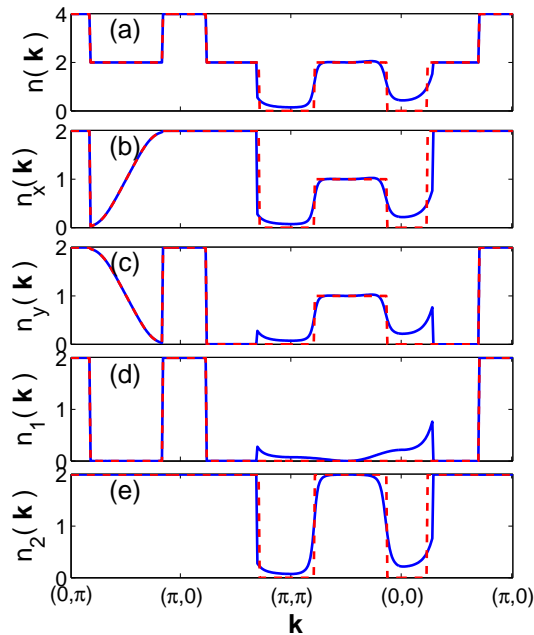}
\caption{(Color online) (a) Total occupation number $n({\bf k})$  for the 
two-orbital model
  Eq.~(\ref{eq:bcs_ours})  with pairing interaction
  $V=0.5$ (continuous lines) and $V=0$ (dashed lines). Parameters: $t_1=1.3$,  
  $t_2=-1$, $t_3=t_4=-0.85$, $\mu=1.54$. (b) Same as (a) but for the orbital 
$d_{xz}$. (c) Same as (a) but for the orbital $d_{yz}$. (d) Same as
(a) but for band 1. (e) Same as (a) but for band 2.\label{fig:nkxy05}}
\end{figure}

\subsection{Non-interacting limit}

\begin{figure}
\includegraphics[width=0.4\textwidth]
{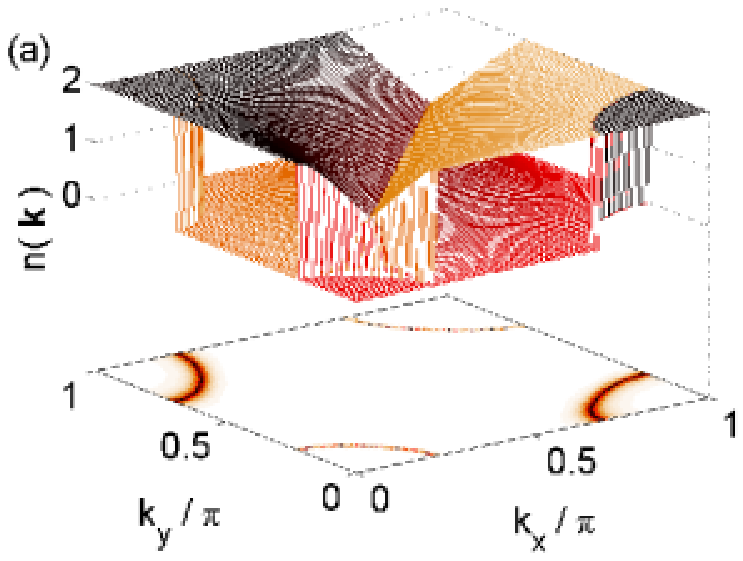}
\includegraphics[width=0.4\textwidth]
{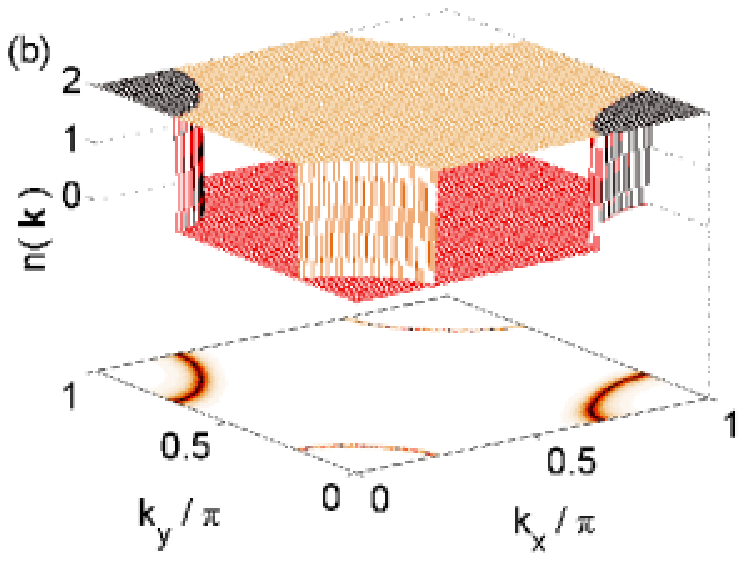}
\caption{(Color online) (a) Occupation number $n({\bf k})$  for the 
two-orbital model
  Eq.~(\ref{eq:bcs_ours})  without pairing interaction for the orbital 
$d_{xz}$ (orange/light) and the orbital $d_{yz}$ (red/dark). The `floor'
indicates the FS in red. (b) Same as (a) but for band 1 (orange/light) and band 2 
(red/dark)\label{nkxysurV0}}
\end{figure}

Let us start with the non-interacting case in which $V=0$. The spectral 
functions along high-symmetry directions in momentum space are presented in 
Fig.~\ref{fig:Akb2g_00} and, as expected, they reproduce the non-interacting 
band dispersion. We also show the total occupation number $n({\bf k})$ along the
same directions [dashed lines in 
Fig.~\ref{fig:nkxy05}(a)], as well as the occupation number for each orbital 
$n_{x}({\bf k})$ [dashed lines in panel (b)] and $n_{y}({\bf k})$ 
[dashed lines in panel (c)] and
the orbital occupation in the quadrant of the first BZ defined by 
$0\le k_x, k_y\le\pi$ in Fig.~\ref{nkxysurV0}(a).
It is clear that the electron-like and hole-like Fermi surfaces are determined by an 
admixture of the two orbitals. 

On the other hand, in the band representation, band 1 determines the electron 
pockets while band 2 forms the hole pockets as it can be seen from the 
behavior of
$n_1({\bf k})$ and $n_2({\bf k})$ (indicated by the dashed lines in panels 
(d) and (e) of Fig.~\ref{fig:nkxy05}) and by the light (orange) and
dark (red) surfaces in 
Fig.~\ref{nkxysurV0}(b) where the FS is also indicated. It is clear that the
electronic occupation of band 1 is smaller than the electronic population 
of band 2, 
so that unpaired electrons would be expected to belong
predominantly to band 1,
as in the simple model of the previous Sec.~\ref{sec:interband}.  
It is 
interesting to notice that in the orbital representation, on the other hand, 
the electrons are equally distributed among the $xz$ and $yz$ $3d$ orbitals.

\subsection{Nonzero pairing}

Let us discuss what occurs when the pairing interaction becomes nonzero.
To simplify the discussion, define the following points in momentum space: 
$X=(\pi,0)$, $Y=(0,\pi)$, $\Gamma=(0,0)$, and $M=(\pi,\pi)$.
As remarked in Ref.~\onlinecite{moreo}, the $B_{2g}$ pairing 
operator always
has nodes along the $X-Y$ direction because the spatial form factor  
$f({\bf k})=\cos k_x+\cos k_y$ 
vanishes along that line. But, as soon as $V$ is finite, a 
gap opens along the $\Gamma-M$  direction [notice that along this direction 
the pairing is purely intraband since in Eq.~(\ref{6}), $v^2=u^2$ and thus $V_{B}=0$
in Eq.~(\ref{8})]. Along $\Gamma-X$, $\Gamma-Y$, $X-M$, and $Y-M$ nodes 
associated to the different number of electrons in band 1 and band 2 remain
[notice that along these directions the pairing 
is purely interband since $V_{A}=0$ because Eq.~(\ref{eq:bcs_ours}) 
and Eq.~(\ref{6}) become 
identical to one other]. When the pairing interaction $V$ becomes
strong enough to make $n_1({\bf k})=n_2({\bf k})$, 
as described in the simplified model presented 
in the Sec.~\ref{sec:interband_strong}, these nodes vanish. Along any other direction in the 
BZ a mixture of intra and interorbital pairing will be present.\cite{moreo} 

Let us first consider a relatively small pairing $V=0.5$. In
Fig.~\ref{fig:nkxy05}(a), $n({\bf k})$ is presented along high
symmetry directions. Along $Y-X$ there is no pairing and, thus, $n({\bf k})$ is unchanged 
from the non-interacting case shown in the figure with dashed lines.
Along $X-M$, where only interband pairing 
occurs, no effects are observed at the electron pocket FS but a rounding in 
$n({\bf k})$ indicating pairing is observed at the hole pocket FS. However, 
$n({\bf k})$ shows discontinuities at two points indicating the existence 
of nodes. Along the 
diagonal direction $M-\Gamma$, where all the pairing is intraband, it is found that 
$n({\bf k})$ exhibits standard BCS behavior at both hole Fermi surfaces indicating the
opening of gaps. Finally, along $\Gamma-X$ it can be observed a rounding of $n({\bf k})$ 
at the hole Fermi surfaces, indicating pairing, and a sharp jump at the electron FS. 

We can further analyze the pairing in the orbital representation. The occupation
number for the orbitals $xz$ and $yz$ is shown in 
Figs.~\ref{fig:nkxy05}(b-c). Along $Y-X$, where the pairing is zero, we
observe how the FS for the electron pocket at $Y$ $(X)$ is totally determined by 
electrons in the orbital $xz$ $(yz)$ and how the population of each orbital
varies smoothly between the two Fermi surfaces, 
always satisfying $n_x({\bf k})+n_y({\bf k})=2$. 
From $X$ to $M$,
 $n_x({\bf k})=n_y({\bf k})$ for $k_{Fh}<k<M$ indicating that the electrons at the 
hole pocket FS are paired in a wide region around it, but the pairing region 
is very narrow around the electron pocket because the pairing is reduced by the small value of $f({\bf k})$. Along the diagonal, i.e. 
from $M$ to $\Gamma$, standard intraband pairing occurs at both hole Fermi surfaces and, thus,
$n_x({\bf k})=n_y({\bf k})$ 
while from $\Gamma$ to $X$  $n_x({\bf k})=n_y({\bf k})$ 
for $\Gamma<k<k_{Fh}$ indicating pairing around the hole FS and almost no 
pairing 
occurs at the electron pocket FS. The behavior $n_x({\bf k})=n_y({\bf k})={n({\bf k})\over{2}}$
for all $k$ for which pairing occurs and $n_i({\bf k})$ unchanged from the 
non-interacting value for all other $k$, was observed for all values of $V$. 
For this reason, figures for $n_i({\bf k})$ in the orbital representation will not 
be shown for the additional values of $V$ discussed below.

The population of the different bands is presented in 
Fig.\ref{fig:nkxy05}~(d-e). The Fermi surfaces along $Y-X$ are clearly determined only by 
band 1,
which is the band that forms the electron pockets. It can also be observed 
that there is 
almost negligible pairing at the electron FS along $X-M$, but there is clear 
interband pairing at the hole FS. This is an 
indication that, due to the spatial variation of the pairing interaction, 
the attraction is much stronger at the hole pockets than at the electrons 
pockets. Also notice that at the hole pocket FS the pairing is interband and
thus $n_1({\bf k})=n_2({\bf k})$, but this does not happen along the diagonal direction 
$M-\Gamma$ where intraband pairing occurs, and there are more paired electrons 
belonging to band 2 than to band 1. Along the direction $X-\Gamma$ again we
observed a stronger pairing effect at the hole FS than at the electron one.

\subsection{Spectral functions $A({\bf k},\omega)$}\label{sec:b2g_specs}

In this section, we discuss the form of the spectral functions $A({\bf
  k},\omega)$, which can be measured in angle resolved photoemission spectroscopy (ARPES) experiments, for weak to
strong interorbital pairing. 

\subsubsection{Weak attractive coupling}\label{sec:b2g_weak}

The spectral function for $V=0.5$ is depicted 
in Fig.~\ref{fig:Akb2g_05}. 
As discussed above, the interorbital
$B_{2g}$ operator leads to \emph{intra}-band coupling along the
$\Gamma-M$ line, where one consequently clearly sees the hole pockets
to be gapped.
Along $\Gamma-X$, the pairing is purely \emph{inter}-band, and one 
finds a Fermi surface on both the hole and electron
pockets, indicating that $V=0.5$ corresponds to the ``breached'' phase
in Fig.~\ref{regimes}. 
However, the spectral weight determining 
the hole-pocket node is weak and might thus be missed
in the analysis of experiments. The node on the electron pocket FS, on the other hand, is 
robust and it should be observed, if present. The same occurs along $X-M$; the 
signal for the node at the electron pocket FS should be robust while the one 
at the hole pocket FS will be weak.

\subsubsection{Intermediate attractive coupling}\label{sec:b2g_intermediate}
\begin{figure}
\includegraphics[width=0.4\textwidth]
{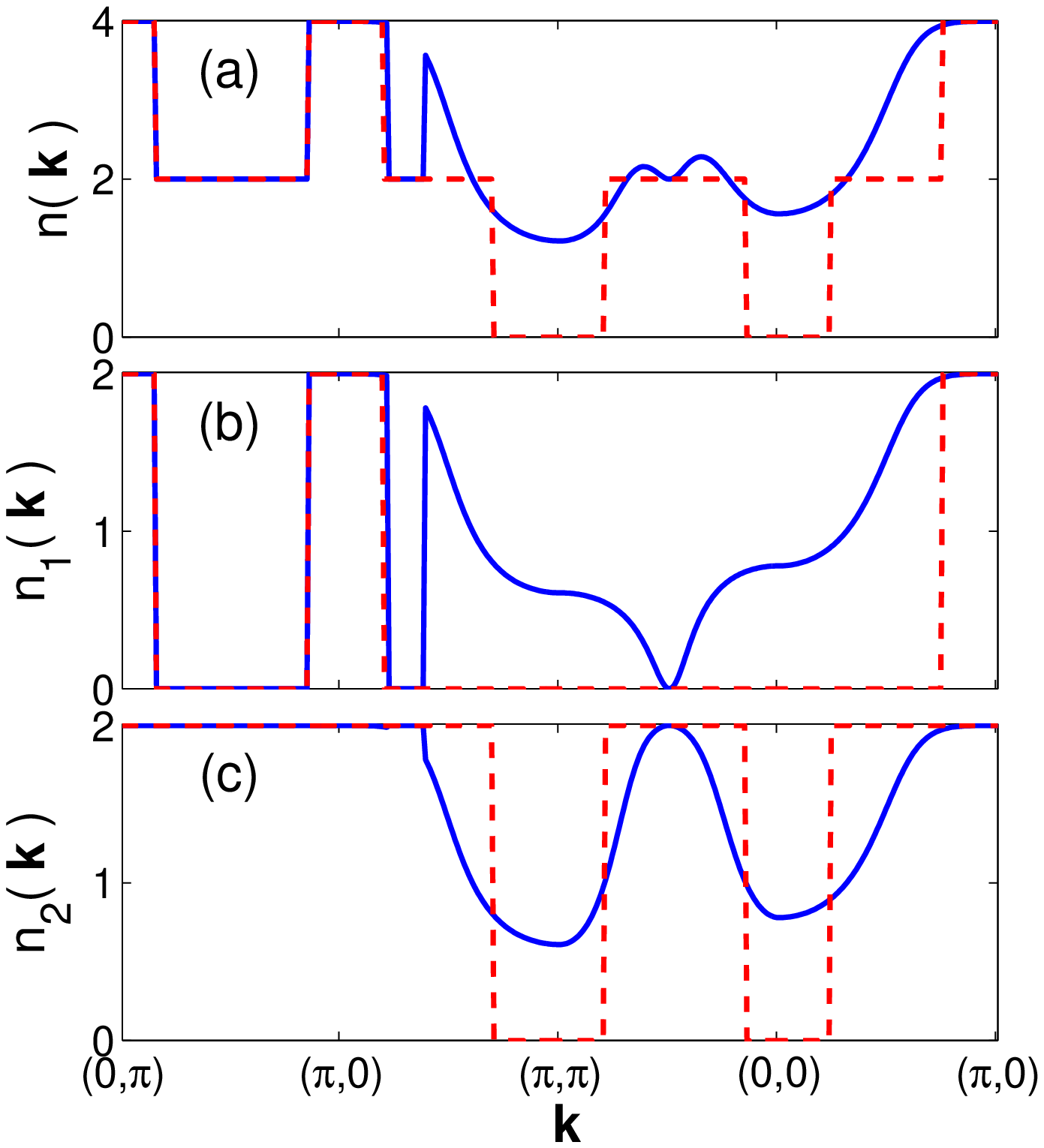}
\includegraphics[width=0.4\textwidth]
{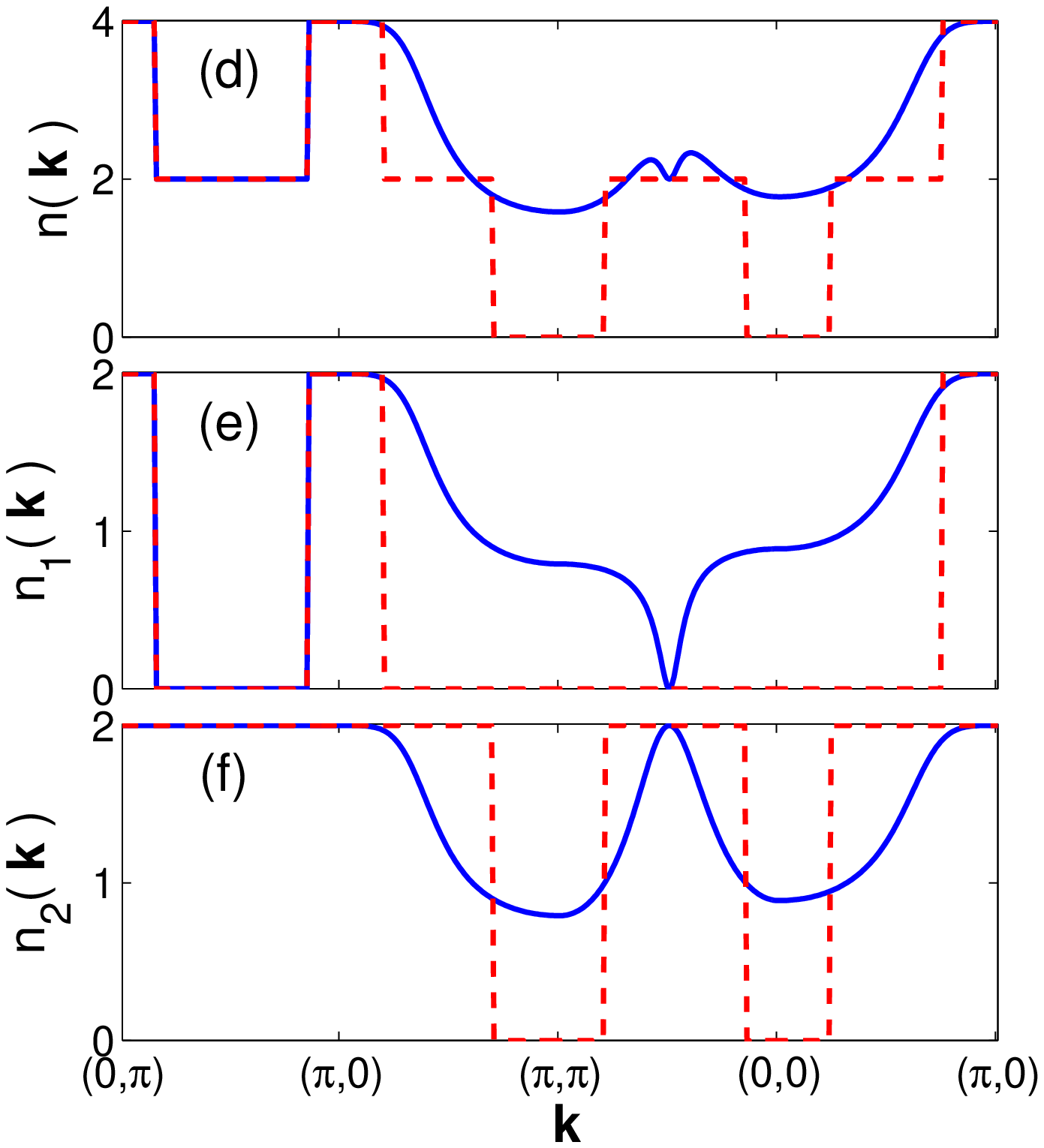}
\caption{(Color online) (a) Total occupation number $n({\bf k})$  for the 
two-orbital model
  Eq.~(\ref{eq:bcs_ours})  with pairing interaction
  $V=3$ (continuous lines) and $V=0$ (dashed lines). Parameters: $t_1=1.3$,  
 $t_2=-1$, $t_3=t_4=-0.85$, $\mu=1.54$. (b) Same as (a) but for band 1. (c) 
Same as (b) but for band 2. (d) Total occupation number $n({\bf k})$  for the 
two-orbital model
  Eq.~(\ref{eq:bcs_ours})  with pairing interaction
  $V=6$ (continuous lines) and $V=0$ (dashed lines). (e) Same as (d) but for 
band 1. (f) 
Same as (e) but for band 2.\label{fig:nk36}}
\end{figure}

As the pairing interaction increases, the nodes resulting from the interband 
pairing should get closer to each other, as discussed in
Sec.~\ref{sec:interband}.
Figures~\ref{fig:nk36}(a-c) show $n({\bf k})$ for $V=3$
along high-symmetry directions in the band representation. One can see
that there are still unpaired electrons, and $V=3$ consequently falls
into the ``breached'' region schematically represented in 
Fig.~\ref{regimes}. 

\begin{figure}
\subfigure{\includegraphics[width=0.4\textwidth]{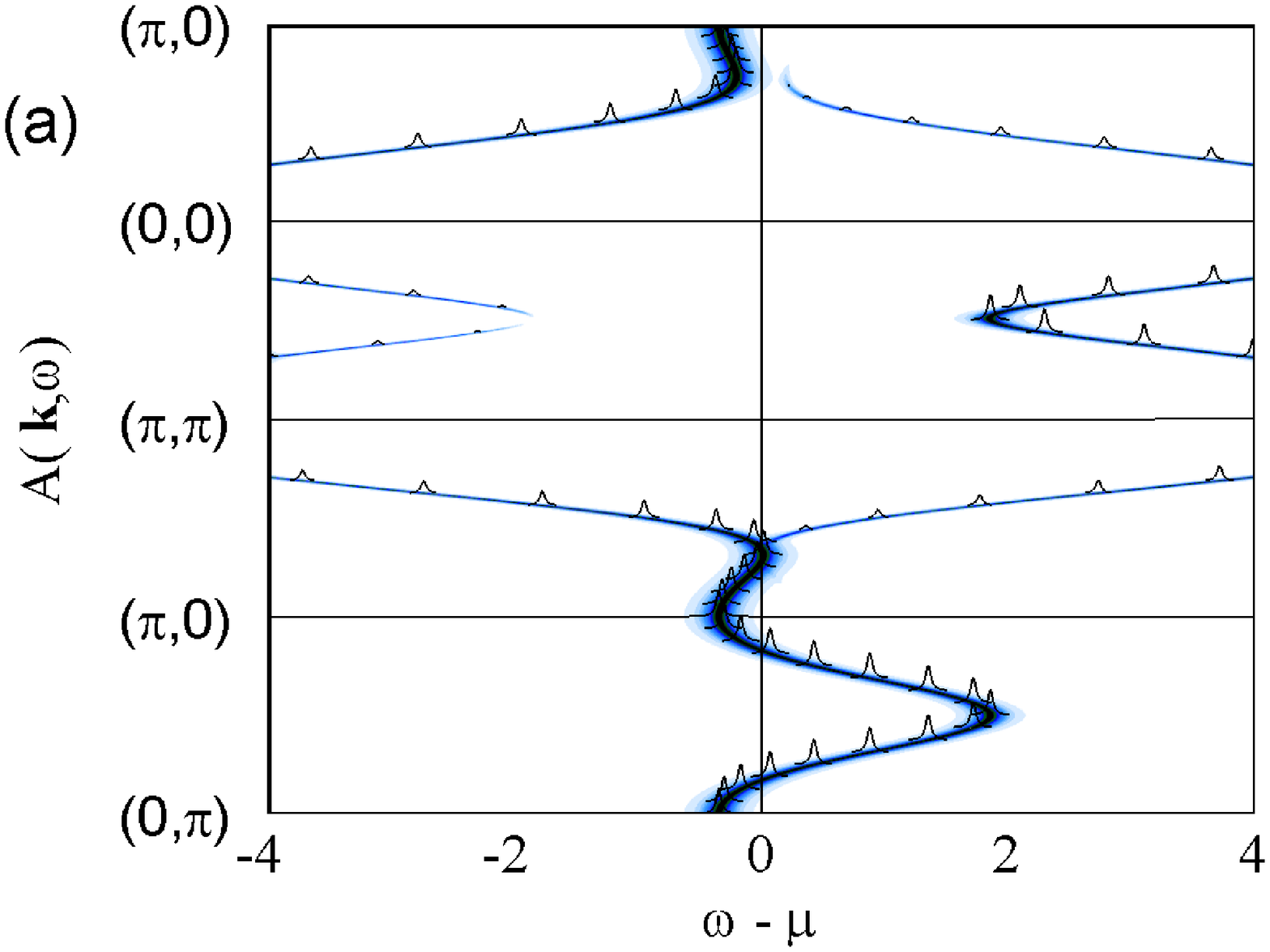}\label{fig:Akb2g_3}}
\subfigure{\includegraphics[width=0.4\textwidth]{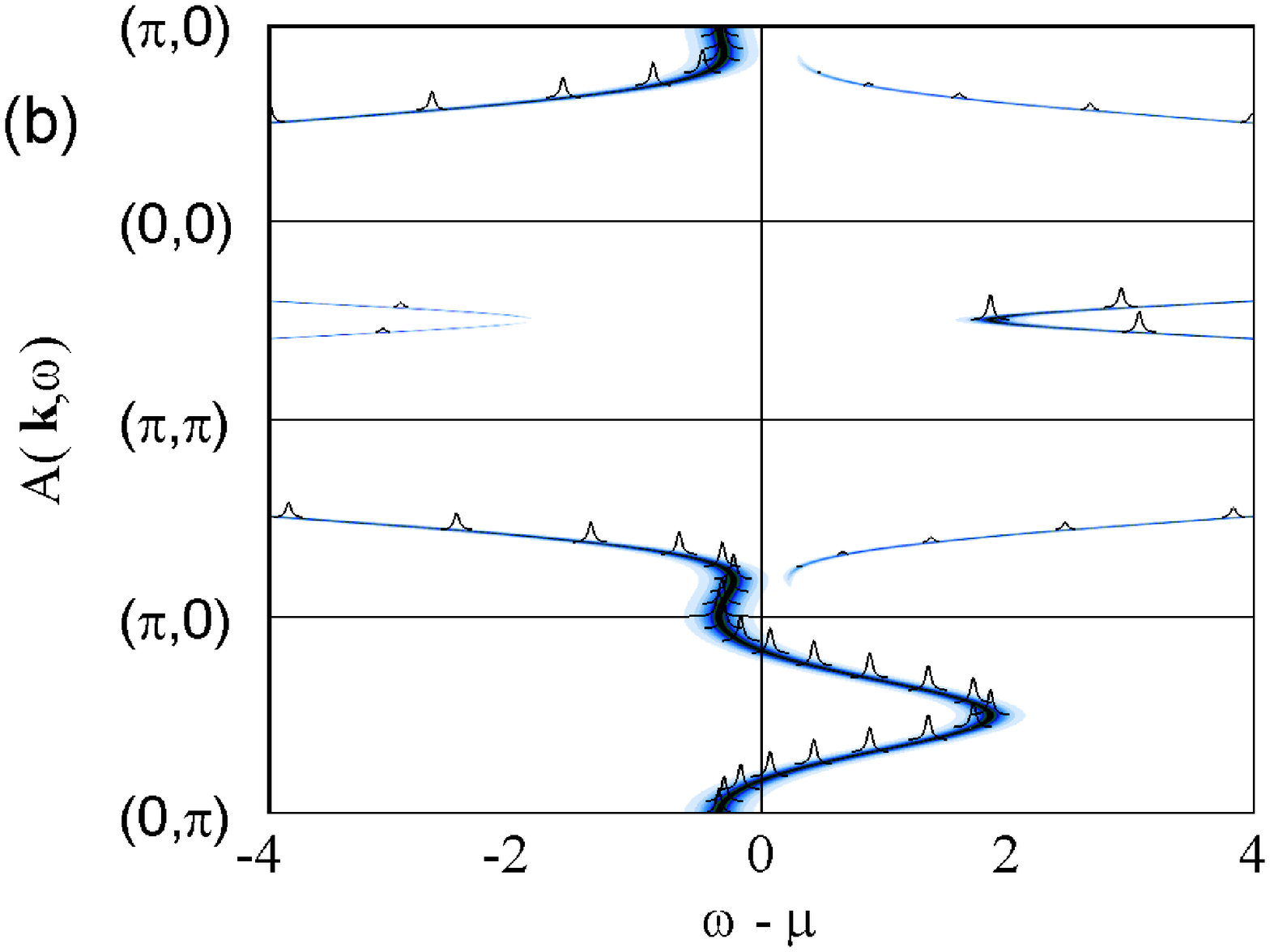}\label{fig:Akb2g_6}}
\caption{ (Color online) (a)Spectral density for the two-orbital model
  Eq.~(\ref{eq:bcs_ours}) for a pairing strength $V=3$. The rest of the
  parameters are as in Fig.~\ref{fig:Akb2g_0}; (b) same as (a) but 
for strong pairing $V=6$. \label{fig:Akb2g_36}}
\end{figure}

It is interesting to notice that while $k'_{Fh}>k_{Fh}$, on the other hand
$k'_{Fe}\approx k_{Fe}$ indicating that the reconstruction around the hole 
pockets is much larger than around the electron pockets. 
This can also
be observed in Fig.~\ref{fig:nk36}(b), where we observe unpaired
electrons along $X-M$, but not along $\Gamma-X$. The effect occurs in part 
due to the smaller value of $f({\bf k})$ at the electron pockets but also because, 
due to the band dispersions, the price in kinetic energy for interband pairing 
is much larger at the electron pockets than at the hole pockets.
The spectral density in Fig.~\ref{fig:Akb2g_3} further illustrates that
the pairing interaction is more effective along $\Gamma-X$ than  
along $X-M$. Only shadow spectral weight 
crosses the chemical potential along $\Gamma-X$, while strong spectral weight 
crosses the chemical potential twice along $X-M$ and leads to two nodes.

\subsubsection{Strong attractive coupling}\label{sec:b2g_strong}
Finally, let us consider a much stronger value of the pairing, such as $V=6$,
in the BCS region of Fig.~\ref{regimes}, for which the 
only nodes observed are those along $X-Y$, due to the vanishing of the 
pairing operator. Figures~\ref{fig:nk36}(d-f), show that $n({\bf k})$ 
is discontinuous only along $X-Y$, 
while it is smooth along all the other directions indicating pairing. Note that
$n_1({\bf k})=n_2({\bf k})$ along $\Gamma-X$ and $X-M$ where interband pairing occurs,
while along $\Gamma-M$, $n_i({\bf k})$ is smooth but different for each band because 
the pairing is intraband. The behavior of the spectral functions displayed in 
Fig.~\ref{fig:Akb2g_6} shows that spectral weight only crosses the 
chemical 
potential along the $X-Y$ direction. In this situation, in the folded BZ, 
nodes should occur only at the points where the two electron pockets cross with
each other, as indicated in Ref.~\onlinecite{moreo}. In the rest of the BZ 
an anisotropic gap will be observed.

\subsection{Stability of the $B_{2g}$ pairing state}

Finally, let us discuss the important issue of the 
stability of the $B_{2g}$ pairing state. 
It will be assumed, following the notation in Appendix A of 
Ref.~\onlinecite{moreo}, that Eq.~(\ref{eq:bcs_ours}) has arisen from an interorbital
attractive potential of the form
\begin{equation}
V_{\bf k,k'}=V^*(\cos k_x+\cos k_y)(\cos k'_x+\cos k'_y),
\label{vkk}
\end{equation} 
\noindent and that 
\begin{equation}
\Delta^{\dagger}({\bf k})=\Delta({\bf k})=V^*\Delta(\cos k_x+\cos k_y),
\end{equation}
\noindent where
\begin{eqnarray}
\Delta({\bf k})&=&-\sum_{{\bf k'}}V_{\bf k,k'}\langle b_{{\bf k'}}\rangle, \nonumber \\
\Delta^{\dagger}({\bf k})&=&-\sum_{{\bf k'}}V_{\bf k,k'}\langle
b^{\dagger}_{{\bf k'}}\rangle.
\end{eqnarray}

\begin{figure}
\includegraphics[width=0.4\textwidth]
{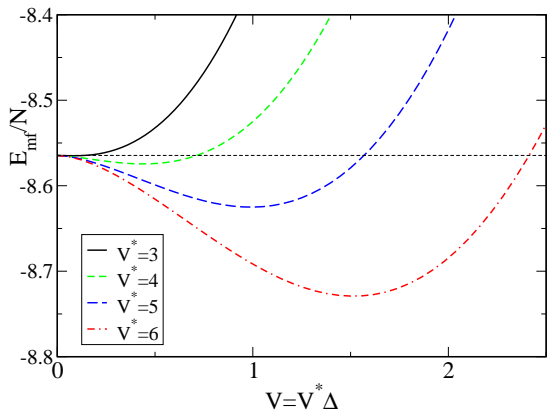}
\caption{(Color online) Mean field energy per unit site for the interorbital
pairing with symmetry $B_{2g}$ for different values of the attraction $V^*$ 
as a function of $V=V^*\Delta$.\label{fig:ebg}}
\end{figure} 

$V$ in Eq.~(\ref{5}) is then given by $V=V^*\Delta$ and the
mean-field energy $E_{\rm MF}$ can be calculated for a given $V^*$ as a function
of $V$. The results are presented in Fig.~\ref{fig:ebg}. In this figure,
a minimum for $V\ne 0$ can be seen for $V^*\ge 3$. This 
indicates that if the pairing attraction overcomes a finite threshold level, the 
mean field results presented here with $V\ge 0.2$ will be stable. 
Thus, \emph{if the attraction exists, both multinodal (breached) and states 
with nodes only along {\rm X-Y} (strong coupling) are possible.}

\section{Conclusions}\label{sec:conclusions}

Summarizing, in this manuscript the possibility of intra and interband pairing 
in multiorbital systems has been discussed. While interband pairing has previously been studied
in the context of QCD, cold atoms,\cite{PhysRevLett.90.047002,PhysRevLett.91.032001} heavy fermions,\cite{khomskii}
cuprates\cite{kheli}, and BCS superconductivity,\cite{kumar} most of the pairing 
operators proposed for the pnictides are based on the premise that the 
pairing has to be purely intraband, i.e., both electrons in the Cooper pairs 
belonging to the same band, compatible with the assumption that in these 
superconductors all the action must occur at the Fermi surfaces.
However, symmetry considerations show that in 
models for the pnictides interorbital pairing is allowed with the two members of the Cooper pair belonging to different bands.\cite{wan} This is not 
surprising since the bands that determine the electron and hole Fermi surfaces consist
of hybridized orbitals. These interorbital pairing operators give rise 
to not only intra but also interband pairing when the band representation is used. In 
addition, numerical calculations in a minimal two orbital model for the 
pnictides favor one of these non-trivial pairing operators.\cite{ours,moreo} 
As a consequence, a clear discussion of the role of interband pairing 
is necessary. 

The explicit 
calculations shown here of the electronic occupation in the orbital and the band 
representations, as well as the calculation of the spectral functions showing 
the distribution of Bogoliubov bands, may offer guidance in the interpretation of 
ARPES experiments. In particular, most ARPES measurements determine the FS in
the normal state and study the opening of the superconducting gap by monitoring
$A(k_F,\omega)$ as they lower the 
temperature.\cite{arpes,arpes2,arpes22,arpes3,hsieh} 
Notice that this approach would
miss the nodes associated with the ``breached'' phase since in the 
superconducting state a gap would be observed in $A(k_F,\omega)$ while the
node would be detected in $A(k'_F,\omega)$. Thus, experimentalists should
investigate the possibility of nodes at points in momentum space that do not 
belong to the normal state FS and they must keep in mind that some of the 
nodes may be determined by shadow bands with very small spectral weight.

The most recent experimental results with the polarization dependence of the 
ARPES spectra for BaFe$_{1.85}$Co$_{0.15}$As$_2$ provided the allowed contribution
of each of the five $3d$ orbitals to the electron and hole 
Fermi surfaces.\cite{FeAs_orb_FS} While discrepancies with
proposed four and five orbital models are remarked in that publication, it is
interesting to observe that the 
minimal two-orbital model addressed here does not contradict the ARPES findings if
we disregard the additional $\beta$ hole pocket FS that they present, which 
is reasonable because it has $d_{x^2-y^2}$ character, an orbital not 
included in the minimal model considered here. In fact,
along their $\Gamma-M$ direction, which corresponds to our $\Gamma-X$, our first
hole FS is purely $d_{xz}$ as it is their $\alpha_{\pi}$ hole FS; our second
hole FS, which arises upon folding our extended FS along $X-Y$ is purely $d_{yz}$,
as it is their $\alpha_{\sigma}$ hole FS, and our electron FS is purely $d_{yz}$ as
it is their $\gamma$ electron FS. Our second electron FS (obtained upon folding) has a
purely $d_{xz}$ character, while their electron FS $\gamma'/\alpha'$ appears to 
be mostly $d_{yz}$ and $d_{xy}$, but with some amounts of $d_{xz}$ 
as well. Along
the diagonal direction $\Gamma-X$, which corresponds to our $\Gamma-M$, the two
hole Fermi surfaces are a symmetric admixture of $d_{xz}$ and $d_{yz}$, exactly as in the 
two-orbital model.\cite{hoppings} Thus, these similarities between the 
experimental results 
and the band composition of the simple two-orbital model offer encouragement 
towards exploring whether the main physics of the pnictides can be captured 
with such a minimum number of degrees of freedom.

We have also mentioned in the text that the symmetry of the lattice and of the orbitals 
introduce constraints on the possible pairing operators. In general, a purely 
intraband pairing, such as 
the proposed $s_{\pm}$ state,\cite{kuroki,Mazin:2008p1695,korshunov,parker} would 
occur only if the coupling of the electrons with the source of the attraction 
is \emph{identical} for all orbitals. 
Considering the different spatial orientations of the orbitals, it is not 
obvious that this should be the case, since, as discussed in the Introduction, 
phonons couple differently to electrons in the $p_z$ and $p_x$ boron orbitals 
in the case of MgB$_2$. Thus, if indeed the coupling results to be the same for all 
$3d$ 
orbitals we could use this fact to elucidate the pairing mechanism; but, if 
this is not the case, we would expect some degree of interband pairing, at least in some 
regions of the Brillouin zone and, thus, it is important to study the 
experimental and theoretical consequences of such a possibility.

\acknowledgments
This work was supported by the NSF grant DMR-0706020 and the
Division of Materials Science and Engineering, U.S. DOE, under contract
with UT-Battelle, LLC.

%\bibliographystyle{prsty-etal}
%\bibliography{feas}

\begin{thebibliography}{10}

\bibitem{Kamihara:2008p932}
Y. Kamihara, T. Watanabe, M. Hirano, and H. Hosono, J. Am. Chem. Soc. {\bf
  130},  3296  (2008).

\bibitem{chen1}
G.~F. Chen, Z. Li, G. Li, J. Zhou, D. Wu, J. Dong, W.~Z. Hu, P. Zheng, Z.~J.
  Chen, H.~Q. Yuan, J. Singleton, J.~L. Luo, and N.~L. Wang, Phys. Rev. Lett.
  {\bf 101},  057007  (2008).

\bibitem{chen2}
G.~F. Chen, Z. Li, D. Wu, G. Li, W.~Z. Hu, J. Dong, P. Zheng, J.~L. Luo, and
  N.~L. Wang, Phys. Rev. Lett. {\bf 100},  247002  (2008).

\bibitem{wen}
H.-H. Wen, G. Mu, L. Fang, H. Yang, and X. Zhu, Europhys. Lett. {\bf 82},  17009  (2008).

\bibitem{chen3}
X.~H. Chen, T. Wu, G. Wu, R.~H. Liu, H. Chen, and D.~F. Fang, Nature {\bf 453},
   761  (2008).

\bibitem{ren1}
Z. A. Ren, J. Yang, W. Lu, W. Yi, G. Che, X. Dong, L. Sun, and Z. Zhao, Mater.
  Res. Innovat. {\bf 12},  105  (2008).

\bibitem{55}
Z. A. Ren, W. Lu,  J. Yang, W. Yi, X.-L. Shen, Z.-C. Li, G.-C. Che, X.-L. Dong, L.-L. Sun, F. Zhou, and Z.-X. Zhao, Chin. Phys. Lett. {\bf 25}, 2215  (2008).

\bibitem{ren2}
Z.-A. Ren, G.-C. Che, X.-L. Dong, J. Yang, W. Lu, W. Yi, X.-L. Shen, Z.-C. Li,
  L.-L. Sun, F. Zhou, and Z.-X. Zhao, Europhys.~Lett.~{\bf 83},  17002  (2008).

\bibitem{first}
S. Lebegue, Phys. Rev. B {\bf 75},  035110  (2007).

\bibitem{Singh:2008p1736}
D.~J. Singh and M.-H. Du, Phys. Rev. Lett. {\bf 100},  237003  (2008).

\bibitem{xu}
G. Xu, W. Ming, Y. Yao, X. Dai, S.-C. Zhang, and Z. Fang, 
Europhys.~Lett.~{\bf 82},  67002
  (2008).

\bibitem{cao}
C. Cao, P.~J. Hirschfeld, and H.-P. Cheng, Phys. Rev. B {\bf 77},  220506
  (2008).

\bibitem{fang2}
H.-J. Zhang, G. Xu, X. Dai, and Z. Fang, Chin. Phys. Lett. {\bf 26},  017401
  (2009).

\bibitem{plee}
P.~A. Lee and X.-G. Wen, Phys. Rev. B {\bf 78}, 144517 (2008).

\bibitem{Suhl}
H. Suhl, B.~T. Matthias, and L.~R. Walker, Phys. Rev. Lett. {\bf 3},  552
  (1959).

\bibitem{BCS}
J. Bardeen, L.~N. Cooper, and J.~R. Schrieffer, Phys. Rev. {\bf 108},  1175
  (1957).

\bibitem{akimitsu}
J. Nagamatsu, N. Nakagawa, T. Muranaka, Y. Zenitani, and J. Akimitsu, Nature
  {\bf 410},  63  (2001).

\bibitem{Louie}
H.~J. Choi, D. Roundy, H. Sun, M.~L. Cohen, and S.~G. Louie, Nature {\bf 418},
  758  (2002).

\bibitem{Wang2001179}
Y. Wang, T. Plackowski, and A. Junod, Physica C {\bf 355},  179   (2001).

\bibitem{PhysRevLett.87.047001}
F. Bouquet, R.~A. Fisher, N.~E. Phillips, D.~G. Hinks, and J.~D. Jorgensen,
  Phys. Rev. Lett. {\bf 87},  047001  (2001).

\bibitem{PhysRevLett.87.167003}
H.~D. Yang, J.-Y. Lin, H.~H. Li, F.~H. Hsu, C.~J. Liu, S.-C. Li, R.-C. Yu, and
  C.-Q. Jin, Phys. Rev. Lett. {\bf 87},  167003  (2001).

\bibitem{PhysRevLett.87.137005}
P. Szab\'o, P. Samuely, J. Ka\v{c}mar\v{c}\'ik, T. Klein, J. Marcus, D.
  Fruchart, S. Miraglia, C. Marcenat, and A.~G.~M. Jansen, Phys. Rev. Lett.
  {\bf 87},  137005  (2001).

\bibitem{PhysRevLett.87.177008}
F. Giubileo, D. Roditchev, W. Sacks, R. Lamy, D.~X. Thanh, J. Klein, S.
  Miraglia, D. Fruchart, J. Marcus, and P. Monod, Phys. Rev. Lett. {\bf 87},
  177008  (2001).

\bibitem{PhysRevLett.87.157002}
X.~K. Chen, M.~J. Konstantinovi\'{c}, J.~C. Irwin,
  D.~D. Lawrie, and J.~P. Franck, Phys. Rev. Lett. {\bf 87},  157002  (2001).

\bibitem{PhysRevLett.87.177006}
S. Tsuda, T. Yokoya, T. Kiss, Y. Takano, K. Togano, H. Kito, H. Ihara, and S.
  Shin, Phys. Rev. Lett. {\bf 87},  177006  (2001).

\bibitem{kuroki}
K. Kuroki, S. Onari, R. Arita, H. Usui, Y. Tanaka, H. Kontani, and H. Aoki,
  Phys. Rev. Lett. {\bf 101},  087004  (2008).

\bibitem{Mazin:2008p1695}
I. Mazin, D. Singh, M. Johannes, and M. Du, Phys. Rev. Lett. {\bf 101},  057003
   (2008).

\bibitem{FCZhang}
X. Dai, Z. Fang, Y. Zhou, and F.-C. Zhang, Phys. Rev. Lett. {\bf 101},  057008
  (2008).

\bibitem{han}
Q. Han, Y. Chen, and Z.~D. Wang, Europhys.~Lett.~{\bf 82},  37007  (2008).

\bibitem{korshunov}
M.~M. Korshunov and I. Eremin, Phys. Rev. B {\bf 78},  140509  (2008).

\bibitem{Baskaran:2008p832}
G. Baskaran, arXiv:0804.1341, 2008.

\bibitem{yildirim}
T. Yildirim, Phys. Rev. Lett. {\bf 101},  057010  (2008).

\bibitem{Si:2008p1561}
Q. Si and E. Abrahams, Phys. Rev. Lett. {\bf 101},  076401  (2008).

\bibitem{yao}
Z.-J. Yao, J.-X. Li, and Z.~D. Wang, New J. Phys. {\bf 11},  025009  (2009).

\bibitem{xu2}
C. Xu, M. M\"uller, and S. Sachdev, Phys. Rev. B {\bf 78},  020501  (2008).

\bibitem{stratos}
E. Manousakis, J. Ren, S. Meng, and E. Kaxiras, Phys. Rev. B {\bf 78},  205112
  (2008).

\bibitem{dolgov} 
%Notice that some authors use the words 
%``Interband Superconductivity'' to describe processes in which interband 
%hopping of pairs of electrons belonging to the same band is considered. See, 
%for example,
O. Dolgov, I. Mazin and D. Parker, Phys. Rev. B {\bf 79},
060502(R)(2009). 
%This does not describe the situation considered in this work which we 
%call ``Interband pairing''  in which the pairs are formed by electrons in 
%different bands.

\bibitem{ours}
M. Daghofer, A. Moreo, J.~A. Riera, E. Arrigoni, D.~J. Scalapino, and E.
  Dagotto, Phys. Rev. Lett. {\bf 101},  237004  (2008).

\bibitem{scalapino}
S. Raghu, X.-L. Qi, C.-X. Liu, D.~J. Scalapino, and S.-C. Zhang, Phys. Rev. B
  {\bf 77},  220503  (2008).

\bibitem{moreo}
A. Moreo, M. Daghofer, J.~A. Riera, and E. Dagotto, Phys. Rev. B {\bf 79},
  134502  (2009).

\bibitem{PhysRevLett.90.047002}
W.~V. Liu and F. Wilczek, Phys. Rev. Lett. {\bf 90},  047002  (2003).

\bibitem{PhysRevLett.91.032001}
E. Gubankova, W.~V. Liu, and F. Wilczek, Phys. Rev. Lett. {\bf 91},  032001
  (2003).

\bibitem{khomskii}
O. Dolgov, E. Fetisov, D. Khomskii, and K. Svozil, Z. Phys. B {\bf 67},  63
  (1987).

\bibitem{kheli}
J. Tahir-Kheli, Phys. Rev. B {\bf 58},  12307  (1998).

\bibitem{kumar}
N. Kumar and K.~P. Sinha, Phys. Rev. {\bf 174},  482  (1968).

\bibitem{rong} It is interesting to observe that in the undoped limit there are
also three regimes: normal, coexisting magnetism and metallicity, and magnetism 
without a Fermi surface. For details see R. Yu, K. T. Trinh, A. Moreo, 
M. Daghofer, J. A. Riera, S. Haas, and E. Dagotto,  Phys. Rev. B {\bf 79}, 
104510 (2009).

\bibitem{wang}
Z.-H. Wang, H. Tang, Z. Fang, and X. Dai, arXiv:0805.0736, 2008.

\bibitem{shi}
J. Shi, arXiv:0806.0259, 2008.

\bibitem{2orbitals}
W.-L. You, S.-J. Gu, G.-S. Tian, and H.-Q. Lin, Phys. Rev. B {\bf 79}, 014508 (2009)

\bibitem{wan}
Y. Wan and Q.-H. Wang, Europhys. Lett. {\bf 85}, 57007 (2009).

\bibitem{zhou} Y. Zhou, W-Q. Chen, and F-C. Zhang, Phys. Rev. B {\bf 78},
  064514  (2008).

\bibitem{goswami} P. Goswami, P. Nikolic, and Q. Si, arZiv:0905.2634, 2009.

\bibitem{parker}
D. Parker, O.~V. Dolgov, M.~M. Korshunov, A.~A. Golubov, and I.~I. Mazin, Phys.
  Rev. B {\bf 78},  134524  (2008).

\bibitem{arpes}
T. Kondo, A.~F. Santander-Syro, O. Copie, C. Liu, M.~E. Tillman, E.~D. Mun, J.
  Schmalian, S.~L. Bud'ko, M.~A. Tanatar, P.~C. Canfield, and A. Kaminski,
  Phys. Rev. Lett. {\bf 101},  147003  (2008).
 
\bibitem{arpes2}
H. Ding, P. Richard, K. Nakayama, K. Sugawara, T. Arakane, Y. Sekiba, A.
  Takayama, S. Souma, T. Sato, T. Takahashi, Z. Wang, X. Dai, Z. Fang, G.~F.
  Chen, J.~L. Luo, and N.~L. Wang, EPL {\bf 83},  47001  (2008).

\bibitem{arpes22} L. Wray, D. Qian, D. Hsieh, Y. Xia, L. Li, J.G. Checkelsky, A. Pasupathy, K.K. Gomes, C. V. Parker, A.V. Fedorov, G.F. Chen, J.L. Luo, A. Yazdani, N.P. Ong, N.L. Wang, M.Z. Hasan, Phys. Rev. B {\bf 78}, 184508 (2008). 

\bibitem{arpes3} V. B. Zabolotnyy, D. S. Inosov, D. V. Evtushinsky, A. Koitzsch, A. A. Kordyuk, J. T. Park,
D. Haug, V. Hinkov, A. V. Boris, D. L. Sun, G. L. Sun, C. T. Lin, B. Keimer, M.
Knupfer,
B. Buechner, A. Varykhalov, R. Follath, and S. V. Borisenko, Nature
{\bf 457}, 569 (2009).
 
\bibitem{hsieh} D. Hsieh, Y. Xia, L. Wray, D. Qian, K. Gomes, A. Yazdani, G.F.
Chen, J.L. Luo, N.L. Wang, and M.Z.Hasan, arXiv:0812.2289 (2008).
 
\bibitem{FeAs_orb_FS}
Y. Zhang, B. Zhou, F. Chen, J. Wei, M. Xu, L.~X. Yang, C. Fang, W.~F. Tsai,
  G.~H. Cao, Z.~A. Xu, M. Arita, C. Hong, K. Shimada, H. Namatame, M.
  Taniguchi, J.~P. Hu, and D.~L. Feng, arXiv:0904.4022, 2009.

\bibitem{hoppings} Notice that we have exchanged the values of $t_1$ and $t_2$
provided in Ref.~\onlinecite{scalapino} in order to reproduce the ARPES orbital
composition.


\end{thebibliography}

\end{document}